\documentclass[aps,pra,floatfix,nofootinbib,showpacs]{revtex4}
\usepackage{graphicx} \usepackage{amsmath} \usepackage{amssymb}
\newcommand{\BEQ}{\begin{equation}}
\newcommand{\EEQ}{\end{equation}}
\newcommand{\BEA}{\begin{eqnarray}}
\newcommand{\EEA}{\end{eqnarray}}

\renewcommand{\d}{{\rm d}}

\newcommand{\de}{\Delta}

\newcommand{\eps}{\epsilon}

\newcommand{\ssum}{{\sum}}

\newcommand{\dq}{\dot{q}}
\newcommand{\ddq}{\ddot{q}}
\newcommand{\pd}{\partial}
\renewcommand{\a}{\alpha}
\renewcommand{\b}{\beta}
\newcommand{\g}{\gamma}

\newcommand{\comment}[1]{}
\begin{document}
\bibliographystyle{unsrt}

\draft
\title
{Post-adiabatic forces and Lagrangians with higher-order derivatives}
\author{A.E. Allahverdyan$^{\,1)}$ and B. Mehmani$^{\,2)}$
}
\affiliation{$^{1)}$ Yerevan Physics Institute,
Alikhanian Brothers St. 2, Yerevan 375036, Armenia }
\affiliation{$^{2)}$ Institute for Theoretical Physics,
Valckenierstraat 65, 1018 XE Amsterdam, The Netherlands}

\begin{abstract} We study a slow classical system [particle] coupled to
a fast quantum system with discrete energy spectrum. We adiabatically
exclude the quantum system and construct an autonomous dynamics for the
classical particle in successive orders of the small ratio $\eps$ of the
characteristic times in order to uncover new physical phenomena. It is known that in the order $\eps^0$ the
particle gets an additional [Born-Oppenheimer] potential, while in the
order $\eps^1$ it feels an effective magnetic field related to the Berry
phase. In the order $\eps^2$ the motion of the clasical particle can be
reduced to a free [geodesic] motion on a curved Riemannian manifold,
with the metric generated by the excluded quantum system.  This motion
has a number of unusual features, e.g., it combines subspaces of
different (Riemannian and pseudo-Riemannian) signature for the metric
tensor.  In the order $\eps^3$ the motion of the classical particle is
still described by a Lagrangian, but the latter linearly depends on the
particle's acceleration. This implies the existence of a spin tensor
[non-orbital angular momentum] for the particle. This spin tensor is
related to the momentum via an analogue of the zitterbewegung effect.
The Hamiltonian structure of the system is non-trivial and is defined via non-linear Poisson brackets.
The linear dependence of the effective classical Lagrangian on
higher-order derivatives is seen as well in the higher orders $\eps^n$.

\end{abstract}



\maketitle

\section{Introduction}

A recurrent theme in modern physics is to derive autonomous equations of
motion for an open system, i.e., a system that interacts with its
environment \cite{klim,weiss,landau}. Depending on the type of
environment, there are different conditions under which this procedure
is possible. A group of methods, which goes under the name of
system-bath interaction, amounts to isolating a relatively small system
in contact to an equilibrium environment (thermal bath)
\cite{klim,weiss}. The prerequisite of applying the model reduction in
this case is that the reaction of the system on its environment is in a
sense weak \cite{klim,weiss}. One of the main consequences of this
approach is the Langevin equation, which supplements the Newton equation
of motion for the small system by a (random) conservative and and
non-coservative (i.e., non-Lagrangian), velocity-dependent friction
force \cite{klim,weiss}. 

There is another set-up that allows deriving autonomous equations for an
open system. Here the essential condition is that the target system is
much slower than its environment
\cite{landau,essen,robbins,gold,lit,various,birman}. One of the oldest
results in this direction is the Darwin Lagrangian \cite{landau,essen}:
when the characteristic speed $v$ of charges is slow as compared to the
speed $c$ of the electro-magnetic field, the latter can be adiabatically
excluded, producing at the order $(v/c)^2$ the Darwin Lagrangian for the
charges, which has important applications in plasma physics and
astrophysics; see \cite{essen} for a recent review.

In this paper we shall study the quantum-classical (also called
mean-field or hybrid) dynamics, which describes coupled quantum and
classical systems \cite{robbins,gold}. This is the most used set-up for
coupling quantum and classical variables \cite{picard}, and has numerous
applications, e.g., in chemical physics \cite{chem} and in (semi)quantum
gravity \cite{gravity}.  Assuming that the classical system is slow |a
condition that is normally met in practice|we exclude the fast quantum
system and study to which extent the ensuing dynamics of the slow
classical system can be described by an autonomous Lagrangian-generated
equations for the classical coordinates. For example, in the above
Darwin problem, the autonomous dynamics exists up to higher orders in
$(v/c)$, but already at the order $(v/c)^3$ the dynamics of the charges
is not Lagrangian due to the friction (Abraham-Lorentz) force related to
radiation damping, i.e., to emission of electromagnetic field by the
charges \cite{landau}. (In some special cases this friction force is
suppressed and the Darwin Lagrangian can be written including the order
$(v/c)^4$ \cite{landau}). 

For the quantum-classical dynamics it is well known that at the zero
order of $\eps$|where $\eps$ is the small ratio of the characteristic
times for the quantum over the classical system, respectively|the
influence of the quantum system on the classical one can be described by
the Born-Oppenheimer potential energy term
\cite{robbins,gold,lit,picard,chem,gravity}.  It was shown by Berry and
Robbins that in the first order of $\eps$ one gets an effective magnetic
field, which manifests itself as the velocity-dependent term in the
classical Lagrangian \cite{robbins}.  Goldhaber has recently shown that
in the second order $\eps^2$ one gets in the Lagrangian an additional
kinetic energy term, i.e., a quadratic form in slow velocities
\cite{gold}. A very similar result on the order $\eps^2$
was obtained earlier by Weigert and Littlejohn
for two coupled (fast and slow) quantum systems \cite{lit}.

What happens in the next orders?  In particular, how far
we can continue the expansion over $\eps$, still keeping the classical
system Lagrangian? Most importantly, are there new physical effects
essentially related to post-adiabatic corrections? 

Here we answer these questions. It appears that at every order over
$\eps$ one can derive Lagrange equations for the dynamics of the
classical system. However, there is an important difference between the
orders $\eps$ and $\eps^2$ and all successive orders. At the order
$\eps^3$ the classical dynamics is Lagrangian, but the Lagrangian starts
to depend on the higher-order time-derivatives of the classical
coordinates: While the classical Lagrangians normally depend on the
coordinates and their first-order time-derivatives (velocities), at the
order $\eps^3$ we get a Lagrangian that is linear over the classical
accelerations. 

This fact is of conceptual relevance. The classical physics is
essentially based on the Newton's second law that equates acceleration
to the force, which depends only on coordinates and velocities. As a
consequence, the trajectory of the classical motion is fixed via initial
coordinates and initial velocities. In its turn, the Newton's second law
is generated by a Lagrangian, which depends on coordinates and
velocities.  A Lagrangian depending on higher-order derivatives enlarges
the amount of the initial data needed to fix the classical trajectory
and produces equations of motion that go beyond the Newton's law. Such
Lagrangians were phenomenologically introduced at various places and for
various purposes (see, e.g., Ref.~\cite{sample} for a sample of
references), but our result seems to be the first example where a
higher-derivative Lagrangian emerges for an open classical system due to
time-scale separation. We should like to stress that the fast system
being quantum is not important for obtaining the above result. What is
important is that the fast system is integrable, i.e., it admits a full
and globally well-defined set of action-angle variables.

Dependence on higher-order derivatives in the Lagrangian implies a
number of essential changes in the kinematics of the classical system:
the momentum of the classical system depends on the acceleration, while
the full angular momentum tensor is a sum of the usual orbital part and
a term that can be interpreted as the spin of the classical system.  In
the simplest non-trivial case this spin is proportional to the velocity
square of the classical particle.  We show that this implies the
existence of the zitterbewegung effect, where the momentum of the
classical particle (system) is governed by the projected time-derivative
of the spin.  So far the zitterbewegung effect was known only in the
physics of relativistic Dirac electron; see \cite{barut} for a review,
while we show the same effect appears in a purely non-relativistic
slowly evolved classical system due to its coupling to a fast quantum
system. It
appears now that this effect is a part of the physics generated by
higher-order post-adiabatic corrections.  Similar dependence on
higher-order derivatives is expected at higher orders $\eps^n$ with
$n\geq 4$, though in the present paper we restict ourselves with
deriving the effective classical Lagrangian up to the order $\eps^4$. 

While these results concern higher-order (three and more) post-adiabatic
corrections, we found an interesting effect already in the second-order
post-adiabatic correction, which was formally known since
Ref.~\cite{gold,lit}. It appears that the slow classical motion within this
order can be reduced to a free motion on a Riemannian space with a
signature-indefinite metric tensor. This implies a possibility of
interchanging between time-like and space-like coordinates. Recall in
this context that within non-relativistic classical mechanics the
geodesic motion on a curved surface proceeds according to a
positively-defined metric tensor, while the geodesic motion in the
general theory of relativity has a metric tensor with signature
$(1,-1,-1,-1)$ \cite{landau}. In both cases the signature is fixed. 

This paper is organized as follows. In section \ref{q_m} we introduce
the quantum-classical dynamics.  The next section, section
\ref{adiabatics} outlines the adiabatic perturbation theory, which
differs from the standard text-book presentations by a careful
accounting of the higher-order terms.  In section \ref{1+2} we review
the derivation of the classical Lagrangian in the orders $\eps$ and
$\eps^2$. In particular, we reproduce in a systematic way the results
obtained by Berry and Robbins \cite{robbins} and Goldhaber \cite{gold}.
At the order $\eps^2$ this classical Lagrangian corresponds to a
classical particle moving along the geodesics of a curved manifold.  We
calculate the curvature for the simplest non-trivial case and work out
its implications for the stability of the effective classical motion at
the order $\eps^2$. Here we also point out at an unusual scenario
related to the metric of the manifold changing its signature [i.e.,
chaning from a Riemannian to a pseudo-Riemannian manifold]. Section
\ref{+3} derives the classical Lagrangian at the order $\eps^3$, and
shows that the classical Lagrangian in this case depends linearly on the
third-derivative of the classical coordinates. In this section we
explore kinematical consequences of this result and explore its
Hamiltonian description. In section \ref{+4} we deduce the classical
Lagrangian at the order $\eps^4$ and show that it also depends linearly
on higher-order derivatives of the classical coordinates. The last
section presents our conclusions and summarizes the present work.
Several technical question are discussed in appendices. 

\section{Quantum-classical dynamics}
\label{q_m}

Consider a $K$-degree of freedom
classical system with coordinates $q=(q_1,\ldots, q_K)$ and
with Lagrangian
\BEA
\label{00} {\cal L}_0=
\frac{M}{2}\ssum_{\alpha=1}^K \left(\frac{\d q_\alpha}{\d t} \right)^2
-V(q),
\EEA
where $M$ is the mass, and $V(q)=V(q_1,\ldots, q_K)$ is the potential energy.

Now this classical system (or particle) couples to a quantum system with Hamiltonian operator
$H[q(t)]$, which parametrically depends on the classical coordinates.
The quantum system evolves in time according to the Schr\"odinger equation (for simplicity we put
$\hbar=1$)
\BEA
i\pd_t |\Psi\rangle = H[q(t)]|\Psi(t)\rangle,
\label{2p}
\EEA
where $|\Psi\rangle$ is the wave-function, and where $\pd_t=\frac{\pd}{\pd t}$.

The classical part of the dynamics
is written as \cite{picard,chem,gravity} (see \cite{picard} for a derivation)
\BEA
\label{1p}
M \frac{\d^2 q_\mu}{\d t^2}+\pd_\mu V+\langle\Psi(t)|\pd_\mu H[q(t)]|\Psi(t)\rangle=0,\quad \mu=1,\ldots,K,
\EEA
where we defined \footnote{
Note that $\pd_\mu=\pd_{q_\mu(t)}$ acts only
on the coordinates, but not on the velocities, e.g., $\pd_\mu
\dot{q}_\alpha=0$. In particular, $\pd_\mu$ commutes with the total
time-derivative $\frac{\d}{\d s}$.
}:
\BEA
\pd_\mu=\frac{\pd}{\pd q_\mu(t)}.
\EEA

Eq.~(\ref{1p}) is the Newton equation of motion, where besides the
classical force $-\pd_\mu V$, the classical particle experiences an
{\it average} force $-\langle\Psi(t)|\pd_\mu H[q(t)]|\Psi(t)\rangle$
exerted by the quantum systems. In this sense the classical coordinates
play a role of a mean-field \cite{picard}. The main purpose of the present
paper is to understand to which extent this force can be generated by a
Lagrangian which depends on the classical coordinates $q_\alpha$ and
their time-derivatives.

It should be clear from (\ref{2p}, \ref{1p}) that the total average energy is conserved in time:
\BEA
\frac{\d}{\d t}\left(
\frac{M}{2}\ssum_{\alpha=1}^K \left(\frac{\d q_\alpha}{\d t} \right)^2+V(q)
+\langle\Psi(t)|H[q(t)]|\Psi(t)\rangle
\right)=0.
\EEA

We note that the quantum-classical equations of motion (\ref{2p}, \ref{1p}) can be derived from a Lagrangian
\BEA
\label{pakh}
\widetilde{{\cal L}}=\frac{1}{2i}\langle\pd_t \Psi|\Psi\rangle-\frac{1}{2i}\langle\Psi|\pd_t \Psi\rangle
-\langle \Psi|H|\Psi\rangle +\frac{M}{2}\ssum_{\alpha=1}^K \left(\frac{\d q_\alpha}{\d t} \right)^2
-V(q),
\EEA
where as a set of independently varying parameters one should take
$|\Psi\rangle$ and $q$ (or alternatively $\langle \Psi|$ and $q$)
\footnote{As usual, when varying (\ref{pakh}) we put aside the total
time-derivatives, e.g., $\frac{\d}{\d t} \langle\delta
\Psi|\Psi\rangle$.}.  It is seen that $\widetilde{{\cal L}}$ is simply a
sum of the corresponding quantum and classical Lagrangians, which points
out at the naturalness of quantum-classical equations of motion
(\ref{2p}, \ref{1p}). 

Let us briefly comment on derivations of the quantum-classical dynamics
from a full quantum-quantum dynamics.  Such a derivation was carried
out in literature several times; see, e.g.,
\cite{picard,ratner,delos,stenholm}. Moreover, many derivations
of the (semi)classical mechanics from the quantum mechanics can be
adopted for deriving quantum-classical dynamics; see in this context
\cite{heller,bala} in addition to the above references. The main assumption involved in all these
derivations is that fluctuations of classical coordinate(s) are small
\cite{picard,ratner,delos,stenholm,heller,bala}. For deriving the quantum-classical dynamics it is not
necessary that the classical motion as such is slow \footnote{ It is not
excluded that there are situations, where both the classical limit and
adiabatic limit|where the classical motion is slow|are taken
simultaneously. Our consideration does not apply to such situations,
because when looking for post-adiabatic corrections one should
simultaneously account for post-classical corrections, which is
something we do not do. }.  Note that the derivations of the quantum-classical dynamics need not
neglect fluctuations of all pertinent variables, i.e., it need not
impose the full quantum trajectories. It will suffice that the to-be classical sector of the 
dynamics is approximated via suitable Gaussian density matrices
\cite{heller}. Then, the parameters of this matrices satisfy the equations of motion for 
some effective classical systems \cite{heller} \footnote{A general
remarks is in order here. The quantum-classical dynamics is just an approximation which holds under
suitable conditions. There are, however, controversial aspects related to this dynamics, 
which emerged when people wanted to
get a non-perturbative generalization of the mean-field quantum-classical
dynamics; see, e.g., \cite{diosi} for examples. These generalizations are supposed to be closed and self-consistent theories,
where one part of variables is quantum and another is classical. Such
theories (if they exist) would somehow get the same fundamental status
as their limiting cases, i.e., as quantum and classical mechanics. 
Numerous attempts to 
formulate such fundamental quantum-classical theories met with severe
difficulties \cite{diosi}. Those difficulties do not seem to be 
insurmountable, as witnessed by a recent proposal by Hall and Regginato
\cite{hall}. }. In Appendix \ref{apero} we briefly remind the main argument involved in the 
derivation of quantum-classical dynamics. 

\subsection{Classical representation for the quantum system}

Below we are going to concentrate on the adiabatic limit of the
quantum-classical system, where the classical system is slow and the
quantum system is fast, and derive an autonomous equations of motion
for the classical part. A natural question is that why specifically we need the fast
system to be quantum, and would it be possible to obtain the
same result assuming that also the fast system is classical. (Then we would not
need additional conditions for the applicability of the quantum-classical
dynamics, and we could start from the outset with an overally classical
dynamics). 

The answer to this question is that in principle only one feature of the
fast quantum dynamics is needed, that is its integrability [in the sense
of \cite{arnold}]. To support this answer we may note that all the
results of the quantum adiabatics (including the quantum adiabatic
theorem and the Berry phase) have their natural counterparts in the
classical adiabatic theory for integrable system (i.e., the Berry phase
becomes the Hannay angle in the classical theory) \cite{robbins}. 

There is a more direct way of seeing the analogy between the quantum
dynamics and the classical integrable dynamics \cite{chirkov}. The Schr\"odinger
equation (\ref{2p}) can be mapped to a classical dynamics if one
introduces a classical Hamiltonian
\BEA
{\cal H}= \langle \Psi |H|\Psi \rangle ={\sum}_{nm}\kappa_{nm} \Gamma_n^* \Gamma_m,\\
\kappa_{nm}\equiv \langle \gamma_n|H|\gamma_m\rangle, \qquad \Gamma_m\equiv \langle \gamma_m|\Psi\rangle
\EEA
where $\{|\gamma_n\rangle\}$ is some time-independent orthonormal base in the considered Hilbert space. Now introducing the 
angle ($\phi_n$) and action ($I_n$) variables via the phase and modulus of $\Gamma_n$
as $\Gamma_n=\sqrt{I_n}\,e^{i\phi_n}$, one can show that the Schr\"odinger equation for $\Gamma_n$
reduces to the classical Hamilton equation
\BEA
\label{bund}
\dot{I}_n=\frac{\partial{\cal H}}{\partial \phi_n}, \qquad
\dot{\phi}_n=-\frac{\partial{\cal H}}{\partial I_n},
\EEA
This is a classical integrable dynamics, because it has globally well-defined
action-angle variables with trivial Poisson brackets \cite{arnold}. 
Eq.~(\ref{bund}) makes clear that the adiabatic theory for the quantum 
Schr\"odinger equation can be alternatively developed from
the viewpoint of classical integrable systems \cite{chirkov}. 

For definitness, in the present paper we shall confine ourselves with the quantum fast system.

\section{Time-scale separation and adiabatic perturbation theory}
\label{adiabatics}

We shall now assume that there is a time-scale separation: the quantum
system evolves much faster than the classical particle. To make this assumption
more precise and to investigate its consequences, let us recall that the adiabatic
energy levels $\{E_k(q)\}_{k=1}^d$ and the corresponding eigen-vectors
$\{|k(q)\rangle\}_{k=1}^d$ are defined
via the eigen-resolution of the Hamiltonian $H[q]$ at fixed values of
$q=(q_1,\ldots,q_K)$:
\BEA
\label{bekaa}
H[q]|k(q)\rangle=E_k(q)|k(q)\rangle,\qquad
\langle k(q)|l(q) \rangle =\delta_{kl}, \quad k=1,\ldots,d,
\EEA
where $d$ is the total number of energy levels. We shall assume that the
adiabatic energy levels are not degenerate. Then qualitative {\it
sufficient} condition for the time-scale separation is that the
characteristic time of the classical motion is much larger than
$\frac{\hbar}{\Delta }$, where $\Delta $ is the minimal adiabatic energy
gap: $\Delta\equiv {\rm min}_{k\not= l}(|E_k-E_l|)$. \footnote{ This
condition is sufficient, but not necessary for the validity of the
time-scale separation and the consequent adiabatic approach, e.g., the
latter can still hold if certain level-crossings are allowed. We shall
not consider this more general situation in the present paper. }

Note that the adiabatic representation (\ref{bekaa}) has a gauge freedom:
\BEA
\label{golan}
|k(q)\rangle\to
e^{i\alpha_k(q)}|k(q)\rangle,
\EEA
where $\alpha_k(q)$ is an arbitrary
single-values function of $q=(q_1,\ldots,q_N)$. Hence all physical
oservables have to be gauge-invariant.

To reflect mathematically the fact of time-scale separation we shall write
the dependence of the quantum Hamiltonian on the classical coordinates as
\BEA
\label{habana}
H[q_1(\eps t), q_2(\eps t),\ldots],
\EEA
where $\eps$ is a small dimensionless parameter
\BEA
\label{ura}
\eps\ll 1.
\EEA
The time-scale separation, i.e., condition (\ref{ura}), can be generated,
e.g., by a large mass $M$ of the classical particle. Then the classical
particle moves slowly|provided that its initial velocity is small|and
$\eps\sim 1/\sqrt{M}$. This scenario of time-scale separation is
normally met in chemical physics (heavy classical nuclei versus light quantum electrons)
\cite{chem} and semi-quantum gravity \cite{gravity}.

In the Schr\"{o}dinger equation (\ref{2p}) we shall assume that
the initial state $|\Psi(0)\rangle$ is equal to an eigenstate:
\BEA
\label{castro}
|\Psi(0)\rangle=|n(0)\rangle.
\EEA
Within the adiabatic approach
the choice (\ref{castro}) does not imply any serious loss of generality; see  Footnote~\ref{fidelito}.

Our program is now to solve the Schr\"odinger equation (\ref{2p})
under the adiabatic assumption (\ref{habana}, \ref{ura}), and determine, via
this solution, the structure of the averaged force in (\ref{1p}). To
this end, we shall need the adiabatic perturbation theory, which was
developed in \cite{hage,baev}, and which is explained in detail in
Appendix \ref{ap1}. Now we shall recall some basic facts from this theory.
As in any theory that is based on time-scale separation, we should start with
dividing the sought solution into fast and slow components:
\BEA
\label{simon}
|\Psi(t)\rangle &=& e^{-i \int_0^t \d u\, E_n(\eps u)}\, |\psi_n(\eps t)\rangle\\
                &=& e^{-\frac{i}{\eps} \int_0^{s} \d u\, E_n(u)}\, |\psi_n(s)\rangle,
\label{bolivar}
\EEA
where in (\ref{bolivar}) we introduced the slow time $s=\eps t$, and
where $|\psi_n\rangle$ satisfies [see (\ref{2p})]
\BEA
i\eps |\dot \psi_n(s)\rangle=[\, H(s)-E_n(s)]|\psi_n(s)\rangle.
\label{g1}
\EEA
Here dots denote differentiation over the slow time
\BEA
s=\eps t, \qquad
\dot{A}\equiv \frac{\d A}{\d s},
\EEA
and the lower index $n$ in
(\ref{simon}, \ref{bolivar}, \ref{g1}) refers to the initial state
(\ref{castro}). Depending on the context we shall write $H[q(s)]$ as $H(s)$, etc.

Eqs.~(\ref{simon}, \ref{bolivar}) extend to the adiabatic situation the
usual formula for a stationary state of a time-independent quantum
Hamiltonian. This analogy also explains why $|\psi_n\rangle$ in (\ref{simon},
\ref{bolivar}) depends only on the slow coordinate. We see that the
dynamical phase $e^{-\frac{i}{\eps} \int_0^{s} \d u\, E_n(u)}$ is the
fast component of the wave-vector, since due to $\eps\ll 1$ it stronlgy oscillates at slow
times \footnote{\label{fidelito}This fact also explains
why in (\ref{castro}) it suffices to take a single initial wave-vector
and not a superposition of them. Any superposition will bring in the adiabatic limit strong
oscillations for non-diagonal elements of the resulting density matrix.
This will reduce the superposition to the mixture of adiabatic
eigen-vectors, which amounts to studying the consequences of (\ref{castro}), and then
taking the average over the index $n$ with certain time-independent weights.}.

Within the adiabatic perturbation theory, the solution of
(\ref{g1}) can be sought for via expanding over $\eps$ (see
\cite{hage,baev} and Appendix \ref{ap1})
\BEA
\label{g2}
&&|\psi_n(s)\rangle= e^{\int_0^s \d u\, \langle \dot n(u)|n(u)\rangle} |\phi_n(s)\rangle,\\
&&|\phi_n\rangle=|n(s)\rangle+\eps |n_1(s)\rangle+\eps^2 |n_2(s)\rangle+\eps^3 |n_3(s)\rangle+\ldots
\label{g3}
\EEA
The zero order term $|\phi_n\rangle=|n(s)\rangle$ in this expansion is
the statement of the adiabatic theorem.  In (\ref{g2}), $e^{\int_0^s \d
u\, \langle \dot n(u)|n(u)\rangle}$ is the Berry phase factor; it was
separated out for ensuring the proper gauge-covariance \cite{baev}; see
also below.  Note that $\langle \dot n(u)|n(u)\rangle$ is purely
imaginary (due to $\langle n(u)|n(u)\rangle=1$), so that this phase
factor nullifies, if $|n(u)\rangle$ can be chosen to be real.  An
alternative representation of $|\phi_n(s)\rangle$ is
\BEA
\label{kora0}
&&|\phi_n(s)\rangle=\ssum_{k=1}^d c_{kn}(s)|k(s)\rangle, \qquad c_{kn}(0)=\delta_{kn},\\
&& c_{kn}(s)=\delta_{kn}+\eps c^{[1]}_{kn}(s)+\eps^2 c^{[2]}_{kn}(s)+\eps^3 c^{[3]}_{kn}(s)+\ldots
\label{kora1}
\EEA

Let us quote from Appendix \ref{ap1} several basic formulas of the adiabatic perturbation theory:
\BEA
\label{copu1}
&& c^{[1]}_{k \not =n}(s)=\frac{\langle k (s) | \dot n (s)\rangle}{i\Delta_{nk}(s)},
\qquad \Delta_{kl}(s)\equiv E_k(s)-E_l(s), \\
\label{copu2}
&& c^{[1]}_{nn} (s) = -i {\ssum_k}' \int_0^s \d u\, \frac{|\langle k (u) | \dot
n (u)\rangle|^2}{\Delta_{nk}(u)},\\
&& c^{[2]}_{k\not =n}(s) = \frac{i}{\Delta_{nk}}\left[
c^{[1]}_{k\not =n}\,\langle n | \dot n \rangle
-\langle k|\dot n_1\rangle
\right]
\label{bharat}
\EEA
where ${\ssum_{k}}'$ means that $k=n$ is excluded from the summation $\ssum_{k=1}^d$.

Eq.~(\ref{copu1}) for the first-order adiabatic correction is
especially well-known \cite{robbins,gold}. It is certainly less
well-known that the consistent adiabatic perturbation theory generates
another ${\cal O}(\eps)$ term, i.e., $c^{[1]}_{nn}$; see however
\cite{baev}. This term is purely imaginary and it drops out from the
lowest-order post-adiabatic corrections to the averaged force in
(\ref{1p}).

The representation (\ref{kora0}) is gauge-covariant
$|\phi_n(s)\rangle\to e^{i\alpha_n(s)}|\phi_n(s)\rangle$ due to
(\ref{copu1}--\ref{bharat}).  Eq.~(\ref{g2}) is also gauge-covariant
$|\psi_n(s)\rangle\to e^{i\alpha_n(0)}|\psi_n(s)\rangle$, and shows that
the relative phase ${\rm arg}\left[ \langle
n(0)|\phi_n(s)\rangle\right]$ is gauge-invariant. Hence it can be
observed \cite{sjoqvist,mead}. Note that Berry phase factor in
(\ref{g2}) is not gauge-invariant as such, apart of a certain specific
case \cite{sjoqvist,mead,aharonov}.

\subsection{Compact expression for the non-adiabatic force}

Employing (\ref{g1}, \ref{g2}) we now calculate
\BEA
F_{\mu} \equiv \langle \psi_n|\pd_\mu H|\psi_n \rangle& =& \langle \phi_n|\pd_\mu H|\phi_n \rangle \\
&=& \pd_\mu \langle \phi_n| H|\phi_n \rangle -2\Re \langle \pd_\mu \phi_n |H|\phi_n  \rangle \\
&=& \pd_\mu E_n +\pd_\mu [\langle \phi_n| H|\phi_n \rangle-E_n]
+2 \eps \Im \langle \pd_\mu\phi_n|\dot \phi_n\rangle,
\label{g7}
\EEA
where $\Re$ and $\Im$ mean, respectively, the real and imaginary parts.
The factor $\pd_\mu E_n$ is the force generated by the adiabatic
(Born-Oppenheimer) potential $E_n=E_n(q)$. Thus the last two expressions
in (\ref{g7}) represent the non-adiabatic force.  We denote
\BEA
\label{brams}
F_{\mu}=
F^{[0]}_{\mu}+\eps F^{[1]}_{\mu}+\eps^2 F^{[2]}_{\mu}+\eps^3 F^{[3]}_{\mu} + \ldots,
\EEA
where $F^{[0]}_{\mu}=\pd_\mu E_n$. In this context note from (\ref{kora0}) that
\BEA
\label{osiris}
\langle \phi_n| H|\phi_n \rangle-E_n = {\ssum}_{k}'\Delta_{kn}|c_{kn}|^2={\cal O}(\eps^2).
\EEA

\section{First- and second-order post-adiabatic force.}
\label{1+2}

Using (\ref{kora1}) and (\ref{g7}) we get for the first-order post-adiabatic force:
\BEA\label{brat0}
F^{[1]}_{\mu} =
2 \Im \langle \pd_\mu n|\dot n\rangle
=2 \dq_\a \Im \langle \pd_\mu n|\pd_\a n\rangle,
\label{tu1}
\EEA
where we always assume implicit summation from $1$ to $K$
over the repeated Greek (not Latin!) indices:
\BEA
A_\a B_\a\equiv \ssum_{\a=1}^k A_\a B_\a.
\EEA

Since $\Im \langle \pd_\mu n|\pd_\a n\rangle=-\Im
\langle \pd_\a n|\pd_\mu n\rangle$, Eq.~(\ref{tu1}) leads to an
effective Lorentz [or gyroscopic] force \cite{robbins}.

Employing (\ref{kora0}, \ref{copu1}) we obtain at the second order:
\BEA
\label{brat1}
F^{[2]}_{\mu} &=& \pd_\mu {\ssum}_{k}'\Delta_{kn}|c^{[1]}_{kn}|^2 +2\Im\left(
\langle \pd_\mu n_1|\dot n\rangle+\langle \pd_\mu n|\dot n_1\rangle\right)\\
&=&\pd_\mu {\ssum}_{k}'\frac{\langle k|\dot n\rangle\langle \dot n|k\rangle}{\Delta_{kn}}
+2\Im\left( \frac{\d}{\d s}
\langle \pd_\mu n|n_1\rangle+\pd_\mu\langle n_1|\dot n\rangle\right).
\label{brat11}
\EEA
Note that the non-local contribution $c^{[1]}_{nn}$ drops out from (\ref{brat11}),
since both $c^{[1]}_{nn}$ and $\langle n |\pd_\mu n\rangle$ are purely imaginary.

Working out (\ref{brat1}, \ref{brat11}) and
taking the result together with (\ref{brat0}), we obtain
that in the first and second orders the averaged
force can be generated by the following Lagrangian
\footnote{\label{eulagr}Let we are given a classical system with action
$\int_0^S \d s {\cal L}[\,\dot{q}(s), q(s)\,]$,
where ${\cal L}$ is the Lagrangian, $q$ is the vector of (generalized)
coordinates, and $\dot{q}=\frac{\d q}{\d s}$. The Euler-Lagrange variational equations of motion
$\frac{\d}{\d s}\frac{\pd {\cal L} }{\pd \dq_\mu}-\frac{\pd {\cal L} }{\pd q_\mu}=0$,
are obtained when varying the action over the the coordinate-path $q(s)$ assuming
that the end-points are fixed: $\delta q(0)=\delta q(S)=0$. This well-known set-up has
a straightforward generalization for a Lagrangian
${\cal L}[\,\ddot{q}(s),\,\dot{q}(s), q(s)\,]$ that depend on the acceleration [or more generally
on higher-order derivatives of coordinates]. The corresponding
Euler-Lagrange equations of motion read:
$\frac{\d}{\d s}\frac{\pd {\cal L} }{\pd \dq_\mu}-\frac{\pd {\cal L} }{\pd q_\mu}
-\frac{\d^2 }{\d s^2}\left[\frac{\pd {\cal L} }{\pd \ddq_\mu}\right]=0$,
where now we assume that $\delta q(0)=\delta q(S)=\delta \dot q(0)=\delta \dot q(S)=0$.}
\BEA
\eps F^{[1]}_{\mu} +\eps^2 F^{[2]}_{\mu} &=&
\frac{\d}{\d s}\frac{\pd {\cal L}^{[12]} }{\pd \dq_\mu}-\frac{\pd {\cal L}^{[12]} }{\pd q_\mu},
\EEA
\BEA
{\cal L}^{[12]}(\dot{q}, q)
&=& \frac{\eps^2}{2} G_{\alpha\beta}(q) \dot{q}_{\alpha} \dot{q}_\beta+\eps
A_{\alpha}(q) \dot{q}_\alpha,\\
&& A_\alpha= \Im\langle \pd_\a n|n\rangle,\\
&& G_{\a\b}=-2{\ssum_k}'
\frac{1}{\Delta_{nk}}\,
\Re\{
\langle  n|\pd_\b k\rangle  \langle \pd_\a k|n\rangle
\}.
\EEA
Here $A_\a$ is the vector potential that corresponds to the effective
magnetic field (\ref{brat0}) \cite{robbins}, and $G_{\a\b}$ plays  the role of a
coordinate-dependent mass tensor, which is generated from (\ref{brat11})
\cite{gold}. Note that $G_{\a\b}$ is a positive matrix, i.e.,
$G_{\a\b}\phi_\a\phi_\b\geq 0$, for any vector $\phi_\a$, provided that
the the quantum system starts its evolution from the ground state:
$\Delta_{kn}\geq 0$. Now $G_{\a\b}$ cannot be a positive matrix for all
initial states of the quantum system, since, e.g., when $d=2$ (two level
situation), one has $G_{\a\b}[{\rm excited\, state}]=-G_{\a\b}[{\rm
ground\, state}]$.

The complete classical Lagrangian to the second order is obtained by
adding ${\cal L}^{[12]}(\dot{q}, q)$ and the Born-Oppenheimer potential
$E_n(q)$ to the initial (bare) classical Lagrangian ${\cal L}_0$, given
by (\ref{00}):
\BEA
\label{totaal} {\cal L}_2=
\frac{M}{2}\ssum_{\alpha=1}^K \left(\frac{\d q_\alpha}{\d t} \right)^2
-V(q)-E_n(q)+{\cal L}^{[12]}(\dot{q}, q).
\EEA

Note that when the time-scale separation is enforced by a large [bare]
mass $M$ of the classical particle, the post-adiabatic Lagrangian ${\cal
L}^{[12]}(\dot{q}, q)$ is small as compared to the large kinetic energy
$M\ssum_{\alpha=1}^K \left(\frac{\d q_\alpha}{\d t} \right)^2$; to make
this fact explicit, we rescale this kinetic energy to the slow time via $\eps\sim
1/\sqrt{M}$.

\subsection{Second-order post-adiabatic force, metric tensor and curvature.}

The kinetic part $\frac{\eps^2}{2}[M\delta_{\a\b}+ G_{\alpha\beta}(q)]
\dot{q}_{\alpha} \dot{q}_\beta$ of the second-order Lagrangian
(\ref{totaal}) corresponds to a free particle moving on a Riemannian
manifold with metric tensor \cite{landau}:
\BEA
g_{\a\b}(q)\equiv \eps^2 [
M\delta_{\a\b}+ G_{\alpha\beta}(q)].
\label{mahari}
\EEA
There is an important particular case, where the complete Lagrangian
(\ref{totaal}) just reduces to this kinetic energy.  This happens when
{\it i)} the eigenvectors $|n\rangle$ can be chosen real, which then
nullifies the vector potential $A_\a$, {\it ii)} the bare potential
$V(q)$ and the Born-Oppenheimer potential $E_n(q)$ compensate each
other, $V(q)+E_n(q)=0$, e.g., $V(q)$ is zero from the outset, while
$E_n(q)$ turns to zero, since the eigenvalues of the quantum Hamiltonian
$H[q]$ do not depend on the coordinates $q$ (though the eigen-vectors
do).

Thus we focus on the purely kinetic Lagrangian
\BEA
\label{bohemia}
\frac{1}{2}g_{\a\b}(q)
\dot{q}^{\alpha} \dot{q}^\beta.
\EEA
Once we are going to excersise on the Riemannian geometry, we recover
for the velocities the explicitly contravariant notations \cite{landau}
$\dq^\a$. The metric tensor $g_{\a\b}$ is then naturally covariant. The
Lagrangian (\ref{bohemia}) yields the following equations of motion
\BEA
\ddot{q}^{\alpha} + \Gamma^{\alpha}_{\mu \nu}\dot{q}^\mu \dot{q}^\nu = 0.
\EEA
This is the geodesic equation
$\frac{D \dot{q}^\a}{\d s^2}=0$, where the covariant differential of any vector $C^\a$ is defined as
\BEA
D C^\a=\d C^\a+\Gamma^\a_{\nu\mu}C^\nu \d q^\mu,
\EEA
and where the connections $\Gamma^{\alpha}_{\mu \nu}$ relate to the metric
tensor as  \cite{landau}:
\BEQ
\label{benti}
\Gamma^{\alpha}_{\mu \nu}= \frac{1}{2}
g^{\alpha \sigma} \left(\partial_\mu g_{\sigma \nu} + \partial_\nu g_{\sigma \mu}
- \partial_\sigma g_{\mu \nu}\right).
\EEQ
Here $g^{\alpha \sigma}$ is the inverse of the metic tensor: $g^{\alpha \sigma}g_{\sigma \b}
=\delta_\b^\a$, and where $\delta_\b^\a$ is the Kronecker delta-symbol.

The first important question is whether the Riemannian manifold is
curved or not. In the latter case it is possible to bring $g_{\a\b}$ to
a diagonal and coordinate independent form by going to some new
coordinates $q'$. The criterion of this is the Riemannian curvature
tensor $ R^{\mu}_{\nu \alpha \beta}$ \cite{landau}. It will be more convenient to
present the explicit formula for the covariant curvature tensor \cite{landau}
\BEA
R_{\alpha\, \beta\,\gamma\,\delta}=\frac{1}{2}\left[
\partial^2_{\beta\gamma}g_{\alpha\delta} + \partial^2_{\alpha\delta}g_{\beta\gamma}
-\partial^2_{\beta\delta}g_{\alpha\gamma} -\partial^2_{\alpha\gamma}g_{\beta\delta}
\right]+g^{\mu \nu}\left[
\Gamma_{\nu \,\beta\gamma}\Gamma_{\mu\,\alpha\delta} - \Gamma_{\nu\, \beta\delta}\Gamma_{\mu\,\alpha\gamma}
\right],
\label{tensor}
\EEA
where $\Gamma_{\mu\beta\gamma}=g_{\mu\alpha}\Gamma^{\alpha}_{\beta\gamma}$.
Eq.~(\ref{tensor}) implies the following symmetry relations:
\BEA
R_{\alpha\, \beta\,\gamma\,\delta}=-R_{\beta\, \alpha\,\gamma\,\delta}
=-R_{\alpha\, \beta\,\delta\,\gamma}=R_{\gamma\,\delta\,\alpha\, \beta}.
\label{rsht}
\EEA

If the curvature tensor is non zero the manifold is curved.  The
manifold is not curved, if and only if $ R^{\mu}_{\nu \alpha \beta}=0$.
Recall that for any vector $C^\a$, the curvature tensor determines the
non-commutativity degree of the covariant derivatives \cite{landau}:
\BEA
\label{ali}
C^\a_{\,; \b \, ;\,\g}-C^\a_{\,; \g \, ;\,\b}=-C^\sigma R^\a_{\sigma\b\g}, \qquad C^\a_{\,;\b}\equiv \frac{D C^\a}{\pd q^\b}.
\EEA

It is known that the curvature tensor determines the local behaviour of
geodesics with respect to perturbing their initial conditions
\cite{landau}.  Let $x^\a(s,\phi)$ be a family of geodesics, where $s$
is the time, and $\phi$ is a scalar continuous parameter which
distinguishes the members of the family. Thus by the definition of the
geodesic:
\BEA
\label{barka}
\frac{D u^\a}{\d s}=0, \qquad u^\a\equiv \frac{\partial x^\a}{\partial s}.
\EEA
Introduce a vector $v^\a\equiv \frac{\partial x^\a}{\partial \phi}$, which determines the deviation of two geodesics with
slightly perturbed initial conditions. This vector satisfies the following Jacobi-Levi-Civita equation \cite{landau}
\footnote{To derive Eq.~(\ref{jlc}) note that the very definitions of $u^{\a}$ and $v^{\a}$ imply
$v^\b\, \pd_\b u^{\a}= u^\b\, \pd_\b v^{\a}$, which amounts to
$u^{\a}_{\, ; \b}\, v^{\b}=v^{\a}_{\, ; \b}\, u^{\b}$. Now calculate
directly $\frac{D^2 v^\a}{\d s^2}$ recalling (\ref{ali}) and noting that $u^\a_{\, ; \b}u^\b=0$ due to (\ref{barka}).  }:
\BEA
\label{jlc}
\frac{D^2 v^\a}{\d s^2}= R^\a_{\,\,\beta\gamma\delta}u^\b u^\g v^\delta.
\EEA
The vector $v^a$ can be separated into two components
$v^\a=v^\a_{[1]}+v^\a_{[2]}$: one orthogonal to $u^\a$ ($u_\a v^\a_{[1]}=0$) and another one parallel to $u^\a$.
One can check with help of (\ref{rsht}, \ref{barka}) that the orthogonal component $v^\a_{[1]}$
satisfies the same equation (\ref{jlc}), while the parallel component $v^\a_{[2]}$ satisfies the geodesic
equation (\ref{barka}).

Below we calculate the curvature for the simplest example of two classical coordinates $q^1$ and $q^2$.
The fact of having only two coordinates simplifies the formulas for the curvature.
Eqs.~(\ref{rsht}) imply that there is only one independent
component of the [covariant] curvature tensor, which can be chosen to be $R_{1212}$. All other components are either
zero or equal to $\pm R_{1212}$. One can check that now $R_{\alpha\, \beta\,\gamma\,\delta}$ is expressed as
\BEA
\label{bala}
&& R_{\alpha\, \beta\,\gamma\,\delta}=\frac{R}{2}[g_{\alpha\, \gamma}g_{\beta\, \delta}-g_{\alpha\, \delta}g_{\beta\, \gamma}],
\\
&& R=g^{\alpha\gamma} g^{\beta\delta} R_{\alpha\, \beta\,\gamma\,\delta}=\frac{2R_{1212}}{g_{11}g_{22}-g_{12}^2},
\label{bala1}
\EEA
where $R$ is the scalar curvature. The latter thus determines the whole curvature tensor for the present
two-dimensional situation.
Substituting (\ref{bala}) into (\ref{jlc}) and recalling that one can take $u_\a v^\a=0$
in this equation, we get
\BEA
\label{blot}
\frac{D^2 v^\a}{\d s^2}= -\frac{R}{2} v^\a \,  (u_\b u^\b).
\EEA
Note that $u_\b u^\b$ does not depend on $s$; see (\ref{barka}).

We now set to calculate the curvature tensor $ R^{\mu}_{\nu \alpha
\beta}$ for the simplest possible example, where there are only two
classical coordinates $q^1$, $q^2$ and the quantum system has only two energy levels.
For further simplicity we assume that the quantum Hamiltonian is real.
This means that the Hamiltonian is a linear combination of the first and
third Pauli matrices:
\BEQ
\hat{H} =\left( \begin{array}{cc}
q^2 & q^1  \\
q^1 & - q^2  \end{array} \right).
\EEQ
The eigenvalues and eigenvectors of $\hat{H}$ read respectively
\BEA
\label{grunt1}
&&E_+ = \sqrt{({q^1})^2 + ({q^2})^2}\equiv \rho, \,\,\,\,\, (\rho > 0)\\
&&E_- = - \sqrt{({q^1})^2 + ({q^2})^2}\equiv - \rho,\\
&&\vert + \rangle =  \frac{1}{\sqrt{2 \rho}}\left[\begin{array}{ll}
       \frac{q^1}{\sqrt{\rho - q^2}}&\\
       \sqrt{\rho - q^2} \end{array}
 \right],\\
&& \vert - \rangle =      \frac{1}{\sqrt{2 \rho}} \left[\begin{array}{ll}
    \frac{q^1}{\sqrt{\rho + q^2}}&\\
        - \sqrt{\rho + q^2}\end{array}
 \right].
\label{grunt4}
\EEA
It is seen that the adiabatic energies $E_+$ and $E_-$ cross at $\rho=0$.

We shall study in separate the case when the quantum system starts at $t=0$ from its ground state $\vert - \rangle$,
and from the excited state $\vert + \rangle$.

\subsubsection{The ground state.}

The metric reads form (\ref{mahari}) and (\ref{grunt1}--\ref{grunt4}):

\BEA
\label{galois1}
&&g_{11} = \eps^2 \left[ M + \frac{(q^2)^2}{4 \rho^5}\right],
\qquad g_{22} = \eps^2 \left[ M + \frac{(q^1)^2}{4 \rho^5}\right], \\
&&g_{12} = g_{21} = - \eps^2 \left(\frac{q^1 q^2}{4 \rho^5}\right).\\
\label{galois3}
\EEA
The determinant and trace of the metric read
\BEA
\label{galois4}
{\rm det}[g] = \eps^4 M \left( M + \frac{1}{4 \rho^3}\right), \qquad
{\rm tr}[g] = \eps^2 \left( M + \frac{1}{4 \rho^3}\right).
\EEA
It is seen from (\ref{galois1}--\ref{galois4}) that both the determinant
and the trace of $g_{\alpha\beta}$ are positive; thus the eigenvalues
are positive as well. This situation refers to a usual classical
mechanical particle, which is enforced to move on a two-dimensional
surface.
For the scalar curvature we get from (\ref{benti}, \ref{tensor}, \ref{bala1}) and (\ref{galois1}--\ref{galois4})
\BEA
\label{kusht1}
R = -\frac{3 (1 + 16 M \rho^3)}{2 \eps^2 M \rho^2 \left(1 + 4 M \rho^3\right)^2}.
\EEA
$R$ is strictly negative. Returning to (\ref{blot}) we see that since
the metric is positively defined [see (\ref{galois1}--\ref{galois4})]
$u^\a u_\a$ is always non-negative.  Then the negativity of $R$ in
(\ref{kusht1}) implies that the geodesics are unstable with respect to
small perturbation of initial conditions, because (\ref{blot})
corresponds to a harmonic oscillator with an inverted (though
space-dependent) frequency \footnote{Such a local instability leads to
chaos, if the $(q^1,q^2)$-manifold is compact. This is not the case for
the considered situation, though it is presumably not very difficult to
compactify the manifold, keeping the conclusion on the local instability
of geodesics.}. This unstability might have implications for the
validity of the adopted adiabatic assumption, a question which we plan
to study elsewhere. Note as well that $R$ is singular at $\rho=0$, where
the adiabatic energy levels cross.

\subsubsection{The excited initial state.}

Now we assume that the two-level quantum system starts its evolution
from the excited state $\vert + \rangle$. This case leads to more
interesting possibilities, since now the metric reads:
\BEA
\label{romain1}
&&g_{11} = \eps^2 \left[ M - \frac{(q^2)^2}{4 \rho^5}\right]
\qquad g_{22} = \eps^2 \left[ M - \frac{(q^1)^2}{4 \rho^5} \right],\\
&&g_{12} = g_{21} = \eps^2 \frac{q^1 q^2}{4 \rho^5},\\
\label{romain3}
\EEA
Hence the determinate and trace of $g$ read, respectively,
\BEA\label{detgexc}
{\rm det}[g]= \eps^4 M \left[M - \frac{1}{4 \rho^3} \right], \qquad
{\rm tr}[g]= \eps^2 \left[M - \frac{1}{4 \rho^3} \right].
\EEA
Since the metric (\ref{romain1}, \ref{romain3}) relates to (\ref{galois1}, \ref{galois3}) with transformation
$M\to -M$ and $\eps\to i\eps$ ($i^2=-1$), we get for the scalar curvature directly from
(\ref{kusht1})
\BEA
\label{kusht2}
R = \frac{3 (16 M \rho^3-1)}{2 \eps^2 M \rho^2 \left(1 - 4 M \rho^3\right)^2}.
\EEA
When the particle moves sufficiently far from the origin $q^1=q^2=0$
(i.e., when $4M\rho^3 >1$), the metric is positively defined and the
curvature is positive.  According to (\ref{blot}) this means that the
geodesics are not sensitive to perturbations in initial conditions. At
$4M\rho^3 =1$ the metric tensor changes its signature, so that for $4M\rho^3
<1$ it has one positive and one negative eigenvalue. At $4M\rho^3 =1$
the scalar curvature is singular.  We expect that the adiabatic
assumption will become problematic in the vicinity of the singularity,
but the possibility for the particle to ``tunnel'' between subspaces of
different signature is striking and deserves a more detailed
investigation.

Since the metric tensor is not positively defined for $4M\rho^3 <1$,
(\ref{blot}, \ref{kusht2}) show that for $\frac{1}{4}<4M\rho^3 <1$ the
geodesics with initial condition $u_\a u^\a<0$ can become unstable
\footnote{In the General Theory of Relativity $u_\a u^{\a} <0$ is
prohibited by causality; for massive particles $u_\a u^{\a} >0$ and can
be normalized to $u_\a u^{\a}=1$, while for photons $u_\a u^{\a} =0$
\cite{landau}.  However, for the present classical theory with a
well-defined global time $s$ nothing prohibits us to consider the class
of geodesics with $u_\a u^{\a} <0$.  }.

For even smaller values $16M\rho^3 <1$ of $\rho$ the curvature becomes
negative. Now the unstable geodesics have $u_\a u^\a >0$, while those
with $u_\a u^\a <0$ are (at least locally) stable.

It is thus seen that the initial ground versus the excited state of the
quantum system produce rather different dynamic behaviour for the
classical system.

\section{Third-order post-adiabatic force.}
\label{+3}

We now turn to studying the post-adiabatic force at the order
$\eps^3$.  The calculations here are more involved, though their general
pattern|employing the adiabatic perturbation theory and then
reconstructing the effective Lagrangian|remains the same. The calculation details being
presented in Appendix \ref{third}, we shall quote the main result: At the order $\eps^3$
the non-adiabatic force acting on the classical system is still Lagrangian [see Footnote
\ref{eulagr}]:
\BEA
\label{odessa}
\eps^3 {F_\mu^{[3]}}&= &
\left( \frac{\d}{\d s}\frac{\pd}{\pd \dq_\mu}-\frac{\d^2}{\d s^2}\frac{\pd}{\pd \ddq_\mu}
-\frac{\pd}{\pd q_\mu}  \right){\cal L}^{[3]}[q,\dot q,\ddot q],\\
\label{kerch}
{\cal L}^{[3]}[q,\dot q,\ddot q]&=&
\eps^3\left[ f_{\a\b\g}\dq_\a\dq_\b\dq_\g-z_{\a\b}\ddq_\a\dq_\b\right],
\EEA
where ${\cal L}^{[3]}$ stands for the third-order Lagrangian, and where we defined
\BEA
\label{gagri}
f_{\a\b\g}&=&\Im\left[\,
\langle n|\pd_\gamma n\rangle \langle N_\b|N_\a \rangle +\langle \pd_\g N_\b|N_\a \rangle
\,\right],\\
z_{\a\b}&=&\Im\left[\,
\langle N_\b|N_\a \rangle
\,\right],\\
|N_\mu\rangle &\equiv& {\ssum}_k'
\frac{\langle k|\pd_\mu n\rangle}{\Delta_{nk}}\,|k\rangle.
\label{sochi}
\EEA
It is seen that besides the expected third-order polynomial over the
velocities $f_{\a\b\g}\dq_\a\dq_\b\dq_\g$, the third-order Lagrangian
${\cal L}^{[3]}$ contains a linear dependence on the accelerations $\ddq_\a$. 
The corresponding coupling matrix $z_{\a\b}(q)$ has to be
antisymmetric $z_{\a\b}(q)=-z_{\b\a}(q)$, since any term $\phi_{\a\b}\ddq_\a\dq_\b$ with a
symmetric $\phi_{\a\b}=\phi_{\b\a}$, can be reduced (up to a total
differential in time) to a third-order polynomial over the velocities.

The total Lagrangian describing the classical system including the three-order terms will include the previous
order non-adiabatic forces and the bare classical Lagrangian, ${\cal L}_{3}[q,\dot q,\ddot q]
=(\ref{totaal})+{\cal L}^{[3]}[q,\dot q,\ddot q]$, or
\BEA
\label{brecht}
{\cal L}_{3}[q,\dot q,\ddot q]=-V(q)-E_n(q)+\eps
A_{\alpha}(q) \dot{q}_\alpha+
\frac{\eps^2}{2} [M\delta_{\a\b}+ G_{\alpha\beta}(q)]\dq_\a\dq_\b+
\eps^3\left[ f_{\a\b\g}\dq_\a\dq_\b\dq_\g-z_{\a\b}\ddq_\a\dq_\b\right],
\EEA
while the equations of motion it generates is [see Footnote \ref{eulagr}]
\BEA
\label{barbarossa}
\left( \frac{\d}{\d s}\frac{\pd}{\pd \dq_\mu}-\frac{\d^2}{\d s^2}\frac{\pd}{\pd \ddq_\mu}
-\frac{\pd}{\pd q_\mu}  \right){\cal L}_{3}[q,\dot q,\ddot q]=0.
\EEA
These equations of motion contain third-order time-derivatives
$q^{(3)}_{\a}$ of coordinates, i.e., they can be written as 
\BEA
\label{barsuk}
2\eps^3
z_{\a \b}q^{(3)}_{\b}=\varphi_\a(q,\dq,\ddq). 
\EEA
Thus when ${\rm det}[z_{\a\b}]\not =0$|and this is generically the case for even number of classical
coordinates| the third-derivatives can be expressed through $(q,\dq,\ddq)$.
This means that the dynamics described by (\ref{barbarossa}) needs three independent (sets of)
initial conditions at the initial (slow) time $s_{\rm i}$:
\BEA
\label{dandalosh}
(\,q(s_{\rm i}),\, \dq(s_{\rm i}),\, \ddq(s_{\rm i})\,). 
\EEA

For an odd number $K$ of classical coordinates, one has ${\rm
det}[z_{\a\b}]= 0$, since $z_{\a\b}$ is anti-symmetric. Generically, 
the matrix $z_{\a\b}$ will have only one
eigenvalue equal to zero. Let us denote this eigenvector by
$y^{[0]}_\a$, $y^{[0]}_\a z_{\a\b}=0$, and let $y^{[\g]}_\a$ with
$\g=1,\ldots, K-1$ be the eigenvector of $z_{\a\b}$ with non-zero
eigenvalues $\lambda^{[\g]}$.  \footnote{The
construction described around (\ref{suk}, \ref{suksuk}) is conceptually
not very different from its simplest analog: Consider two classical
degrees of freedom with coordinates $x$ and $q$. Let the corresponding
Lagrangian be $\frac{\dot{q}^2}{2}-V(q,x)$. Note that this Lagrangian
does not contain the kinetic energy for the $x$-particle, i.e., the
kinetic energy matrix is degenerate. The Lagrange equations of motion
read: $\ddq=-V'_q(q,x)$ and $V'_x(q,x)=0$. The second equation is a
constraint on admissible values of $q$ and $x$ at any time.  In
particular, it confines their initial values. Now the initial conditions
amount to $q(s_{\rm i})$, $\dq(s_{\rm i})$ and $x(s_{\rm i})$ provided
that the constraint is satisfied. One is not free in choosing the
initial velocity $\dot{x}(s_{\rm i})$. The latter is determined from
differentiating the constraint over time $s$ and taking $s\to s_{\rm
i}$. } Eq.~(\ref{barsuk}) produces
\BEA
\label{suk}
2\eps^3 \lambda^{[\g]} \, y^{[\g]}_\b
q^{(3)}_{\b}&=&y^{[\g]}_\a \varphi_\a(q,\dq,\ddq), \quad {\rm for} \quad \g=1,\ldots,K-1,\\
0&=&y^{[0]}_\a \varphi_\a(q,\dq,\ddq). 
\label{suksuk}
\EEA
Now the initial conditions $(\,q(s_{\rm i}),\, \dq(s_{\rm i}),\,
\ddq(s_{\rm i})\,)$ at the initial time $s_{\rm i}$ cannot be anymore
taken independently from each other, because (\ref{suksuk}) imposes a
constraint on them.  Provided that $(\,q(s_{\rm i}),\, \dq(s_{\rm i}),\,
\ddq(s_{\rm i})\,)$ satisfy this constraint, (\ref{suk}) gives $K-1$
equations for components of $q^{(3)}_{\a}$. Another equation for
components of $q^{(3)}_{\a}$ can be obtained by differentiating
(\ref{suksuk}) over time $t$ and taking $t\to t_{\rm i}$. 

Before closing this discussion on the initial conditions let us note the
following aspect.  The quantum-classical equations (\ref{2p}, \ref{1p})
have the following well-defined initial conditions at the initial moment
$t=0$ of the fast time $t$: $|\Psi(0)\rangle$, $q(0)$ and $\dq(0)$. On
the other hand, as we saw above, the autonomous classical dynamics 
starts to depend on higher-derivatives of the coordinate(s). The reason
of this difference is that the initial conditions of the autonomous
classical dynamics are to be imposed at an initial value $s_{\rm
i}=\epsilon t_{\rm i}$ of the slow time, where $t_{\rm i}$ should
be still sizably larger than $t=0$. The difference between the original
initial conditions of the slow variables and their effective initial
conditions after eliminating the fast variables is known as the initial slip
problem.  It is well recognized in theories dealing with elimination of
fast variables \cite{slips}. There also exist more or less regular
procedures of relating the original initial conditions to
effective ones \cite{slips}.  We shall not dwell into this issue anymore,
because in the present paper we are interested by autonomous classical
dynamics for sufficiently large (fast) times, where the precise relation
with the original initial conditions is not directly relevant.

\subsection{Kinematics.}

The dependence of ${\cal L}_{3}[q,\dot q,\ddot q]$
on accelerations implies conceptual changes in the kinematics
of the classical system, as we now proceed to show.

First we note that the momentum of the classical system is defined via
the response of ${\cal L}_{3}$ to an infinitesimal coordinate shift
$q_\mu\to q_\mu+\delta q_\mu$, where $\delta q_\mu$ does
not depend on time \cite{landau}: 
\BEA
\delta {\cal L}_{3}=\frac{\pd {\cal L}_{3}}{\pd q_\mu} \delta q_\mu=
\delta q_\mu\frac{\d}{\d s}\left[
\frac{\pd {\cal L}_{3}}{\pd \dq_\mu}
-\frac{\d}{\d s}\frac{\pd {\cal L}_{3}}{\pd \ddq_\mu}
\right],
\EEA
where we used (\ref{barbarossa}).
Thus the momentum is defined as
\BEA
\label{zanzibar}
p_\mu= \frac{\pd {\cal L}_{3}}{\pd \dq_\mu}
-\frac{\d}{\d s}\frac{\pd {\cal L}_{3}}{\pd \ddq_\mu},
\EEA
implying that the equations of motion can be written as $\dot{p}_\mu=\frac{\pd {\cal L}_{3}}{\pd q_\mu}$.

If ${\cal L}_{3}$ would not depend on $q_\mu$ (which is generically not the case),
the corresponding momentum $p_\mu$ will be conserved in time. Note that $p_\mu$ consists of the
usual part $\frac{\pd {\cal L}_{3}}{\pd \dq_\mu} $ and the anomalous part
$-\frac{\d}{\d s}\frac{\pd {\cal L}_{3}}{\pd \ddq_\mu}$ that comes solely from the dependence
of the Lagrangian on the acceleration. Using (\ref{brecht}) we get for the momentum
\BEA
p_\mu=\eps A_\mu+\eps^2 [M\dq_\mu+G_{\mu\a}\dq_\a] +3\eps^3 f_{\mu\a\b}^{\rm (sym)}\dq_\a\dq_\b+2\eps^3
z_{\mu\a}\ddq_\a+\eps^3\,[\pd_\g z_{\mu\b}]\,\dq_\g \dq_\b.
\label{dari}
\EEA
where $f_{\a\b\g}^{\rm (sym)}\equiv\frac{1}{6}\ssum_\Pi f_{\Pi[\a\b\g]}$
is the completely symmetrized expression (\ref{gagri}); the sum is taken
over all six permutations $\Pi$ of three elements. It is seen that the
expression for the momentum does depend linearly on the acceleration.
One half of the acceleration-dependence comes from usual part $\frac{\pd
{\cal L}_{3}}{\pd \dq_\mu} $, while another half comes through the
anomalous part $-\frac{\d}{\d s}\frac{\pd {\cal L}_{3}}{\pd \ddq_\mu}$,
resulting altogether in $2\eps^3 z_{\mu\a}\ddq_\a$ in (\ref{dari}).

The energy corresponding to the Lagrangian ${\cal L}_{3}[q,\dot q,\ddot q]$
is obtained via looking at the total time-derivative of ${\cal L}_{3}[q,\dot q,\ddot q]$:
\BEA
\frac{\d}{\d t}{\cal L}_{3}[q,\dot q,\ddot q]=
\frac{\pd {\cal L}_{3}}{\pd q_\mu} \dq_\mu
+\frac{\pd {\cal L}_{3}}{\pd \dq_\mu} \ddq_\mu
+\frac{\pd {\cal L}_{3}}{\pd \ddq_\mu} \frac{\d^3 q_\mu }{\d s^3},
\label{brut}
\EEA
where we noted that ${\cal L}_{3}[q,\dot q,\ddot q]$ does not have any explicit time-dependence.
Employing equations of motion $\dot{p}_\mu=\frac{\pd {\cal L}_{3}}{\pd q_\mu}$, (\ref{brut}) results into
energy conservation:
\BEA
\frac{\d E}{\d t}=0,\qquad E\equiv
p_\mu\dq_\mu+\frac{\pd {\cal L}_{3}}{\pd \ddq_\mu} \ddq_\mu
-{\cal L}_{3}.
\EEA
For our case the energy $E$ reads
\BEA
E=\frac{\eps^2}{2}[M\delta_{\a\b}+G_{\a\b}]\dq_\a\dq_\b+2\eps^3 f_{\a\b\g}\dq_\a\dq_\b\dq_\g
+2\eps^3 z_{\mu\a}\ddot{q}_\a \dot{q}_\mu
+V(q)+E_n(q).
\EEA
Note that the vector-potential $A_\a(q)$ expectedly drops out
from the expression of energy \cite{landau}. However, the acceleration-dependent part of the Lagrangian
does contribute directly into the energy. In fact, the whole third-order Lagrangian (\ref{kerch}) is multiplied by
2 and enters into the energy.

Let us now turn to the angular momentum tensor, which is defined
via the response of ${\cal L}_{3}$ to an infinitesimal rotation
(i.e., a distance conserving linear transformation) \cite{landau}:
$q_\mu\to q_\mu+\omega_{\mu\sigma}\delta q_\sigma$, where
$\omega_{\mu\sigma}=-\omega_{\sigma\mu}$:
\BEA
\delta {\cal L}_{3}=\omega_{\a\b}
\left[
\frac{\pd {\cal L}_{3}}{\pd q_\a} q_\b
+\frac{\pd {\cal L}_{3}}{\pd \dq_\a} \dq_\b
+\frac{\pd {\cal L}_{3}}{\pd \ddq_\a} \ddq_\b,
\right]=\omega_{\a\b}
\frac{\d}{\d t}\left[
p_\a q_\b+\frac{\pd {\cal L}_{3}}{\pd \ddq_\a} \dq_\b
\right],
\EEA
where we again used (\ref{barbarossa}).
The full momentum tensor is now defined as
[recalling $\omega_{\mu\sigma}=-\omega_{\sigma\mu}$]:
\BEA
M_{\a\b}&=&p_\a q_\b - p_\b q_\a
+\frac{\pd {\cal L}_{3}}{\pd \ddq_\a} \dq_\b
-\frac{\pd {\cal L}_{3}}{\pd \ddq_\b} \dq_\a\\
&=&L_{\a\b}+S_{\a\b},
\EEA
so that when ${\cal L}_{3}$ is rotationally invariant, $M_{\a\b}$ is
conserved in time.  One part of this tensor is the usual orbital
momentum $L_{\a\b}=p_\a q_\b - p_\b q_\a$. The remainder|non-orbital
momentum, or spin| arises due to the dependence of the Lagrangian on the
accelerations, and it is a second-order polynomial over the velocities:
\BEA
\label{uganda}
S_{\a\b}&=&
\frac{\pd {\cal L}_{3}}{\pd \ddq_\a} \dq_\b
-\frac{\pd {\cal L}_{3}}{\pd \ddq_\b} \dq_\a \\
&=& \eps^3 [\,
z_{\b\g}\dq_{\g}\dq_\a-z_{\a\g}\dq_{\g}\dq_\b
\,].
\label{burundi}
\EEA
In the simplest two-coordinate situation $S_{12}=-S_{21}=\eps^3
z_{21}(\dq_1^2+\dq_2^2)$, which means that the spin tensor is
proportional to the velocity square.

\subsubsection{Zitterbewegung effect.}

Now note from (\ref{zanzibar}, \ref{uganda}, \ref{burundi}) and from $z_{\b\a}=-z_{\a\b}$ that the momentum can be written as
\BEA
\label{zi}
p_\mu = \frac{\pd {\cal L}_{3}}{\pd \dq_\mu} +\frac{\d}{\d t} \left[\frac{S_{\a\mu} \dq_{\a}}{\dq^2}\right],
\qquad \dq^2\equiv\dq_\b\dq_\b,
\EEA
which means that the anomalous part $p_\mu - \frac{\pd {\cal L}_{3}}{\pd \dq_\mu}$
of the momentum is driven by the time-derivative of the velocity-projected spin-tensor.

An expression similar to (\ref{zi})|relating the momentum to the
projected time-derivative of the spin|appears in the (relativistic)
Dirac electron theory; see \cite{barut} for a review. There the fact
that the total angular momentum is a sum of the orbital part and the
spin part, as well as the fact that the velocity and the momementum are
different objects and are not simply proportional to each other via the
mass, are the consequence of the relativistic invariance for the
electron. The very effect of the spin time-derivative contributing into
the momentum was named {\it zitterbewegung}, since for the free Dirac
electron this contribution brings in an additional oscilatory motion
\cite{barut}. In a more recent literature, the zitterbewegung effect is
also derived via Lagrangians contaning the higher-order derivatives of
coordinates \cite{salesi,rivas}.

There are, however, several aspects that distinguish (\ref{zi}) from the
zitterbewegung effects already known in literature.  First, we do not
have a relativistic invariant theory; for us relation (\ref{zi}) emerges
due to the fact that the classical system is open. Second, we do not
have to have the conservation of momentum and of angular momentum for
deriving (\ref{zi}).  Both these quantities are generically
non-conserved in our situation (ultimately since the system is open),
but relation (\ref{zi}) still holds generally due to the specific,
anti-symmetric form (\ref{kerch}) of the acceleration-dependent part of
the Lagrangian.

We close this part by re-iterating its main findings: due to interaction
with the fast quantum system the classical system gets a spin
[non-orbital angular momentum], which is related to its momentum via
zitterbewegung effect.

\subsection{Hamiltonian description.}
\label{hamoo}

Further insight into the structure of the effective classical dynamics
is gained by studying its Hamiltonian description.  Within the order
$\eps^2$ the Hamiltonian description is straightforward. However, the
third-order dynamics has a non-trivial Hamiltonian structure, as seen
below.  Rewrite (\ref{brecht}) as
\BEA
{\cal L}_3(q,\dq,\ddq) = L_3(q,\dq)
-\eps^3 z_{\a\b}\ddq_\a\dq_\b,
\label{muskus}
\EEA
where the higher-derivative term is explicitly separated out. Instead of (\ref{muskus})
we now introduce the following extended Lagrangian:
\BEA
\label{batur}
L[q,v,\pi]=\pi_\a (\dq_\a-v_a)-\eps^3 z_{\a\b}\dot{v}_\a v_\b +L_3(q,v) ,
\EEA
which is a function of three set of variables: $q=(q_1,\ldots,q_K)$,
$v=(v_1,\ldots,v_K)$ and $\pi=(\pi_1,\ldots,\pi_K)$. It should be clear
that if we treat $q$, $v$ and $\pi$ as coordinates, then the Lagrange
equations generated by $L[q,v,\pi]$ are equivalent to those generated by
${\cal L}_3(q,\dq,\ddq)$. At this point $\pi$ is considered as a part of the overall set of coordinates. It may be equivalently
viewed as Lagrange multipliers.
If $L[q,v,\pi]$ were not depend on
$\dot{v}$|that is ${\cal L}_3(q,\dq,\ddq)$ were not depend on $\ddq$|we
would write the velocities $v=v(q,\pi)$ as functions of the coordinates
and momenta, and end up with the usual Hamiltonian description with $q$
and $\pi$ being the canonical coordinates and momenta, respectively. Though
$L[q,v,\pi]$ does depend on
$\dot{v}$, it can be still Hamiltonized following to the method advocated in \cite{ja}.

Once the triple $q,v,\pi$ is considered as coordinates, we introduce a separate notation for it
\BEA
Q=(Q_1,\ldots,Q_{3K})=(q_\a,v_\a,\pi_\a).
\EEA
Now the Lagrangian (\ref{batur}) reads
\BEA
\label{jagataj}
L[Q]= A_a[Q] \dot{Q}_a+{\cal H}[Q],\qquad {\cal H}\equiv L_3[Q],
\EEA
where the index $a$ runs from $1$ to $3K$, and where $A_a=(\pi_\a, \eps^3 z_{\b\a}v_\b,0)$ is deduced from (\ref{batur}).
As we show below, ${\cal H}[Q]$ will play the role of Hamiltonian.
Eq.~(\ref{jagataj}) generates the following Lagrange equations of motion:
\BEA
\label{buj}
&& \Omega_{ab}(Q) \,\dot{Q}_b= \frac{\partial {\cal H}}{\partial Q_a}, \\
&& \Omega_{ab}(Q)\equiv\frac{\partial A_a}{\partial Q_b}-\frac{\partial A_b}{\partial Q_a}.
\label{buk}
\EEA
In block-matrix notations $\Omega$ reads
\BEA
\label{avgust}
\Omega = \left( \begin{array}{ccc}
   0 & Y & I \\
-Y^T & Z & 0 \\
  -I & 0 & 0 \\
\end{array} \right),
\EEA
where each element in the above matrix is $K\times K$ matrix:
\BEA
Y_{\a\b}= \eps^3 v_\g \partial_\a z_{\g\b}, \qquad Z_{\a\b}= -2 \eps^3 z_{\a\b}, \qquad I_{\a\b}=\delta_{\a\b},
\EEA
and where $I$ is the $K\times K$ unit matrix. Provided $Z$ is invertible, the inverse of $\Omega$ reads
[block-matrix notations]
\BEA
\Omega^{-1} = \left( \begin{array}{ccc}
   0 & 0 ~&~ -I \\
   0 & Z^{-1} ~&~ -Z^{-1}Y^T \\
   I & -YZ^{-1} ~&~ YZ^{-1}Y^T \\
\end{array} \right).
\EEA
For an even $K$ the matrix $Z$ is generically invertible; compare with our discussion after (\ref{barbarossa}). Since
$\Omega_{ab}$ is invertible, antisymmetric, closed \footnote{Closed
means that $\frac{\partial}{\partial Q_c}\Omega_{ab}
+\frac{\partial}{\partial Q_b}\Omega_{ca}+\frac{\partial}{\partial
Q_a}\Omega_{bc}=0$. This feature is automatic from the definition
(\ref{buk}), and it ensures that the Poisson brackets defined via
$\Omega_{ab}$ does not change in time \cite{arnold}. } and it satisfies
(\ref{buj}), $\Omega_{ab}$ defines a symplectic structure \cite{arnold}. Then ${\cal H}[Q]$ plays the role of Hamiltonian.
These two ingredients are necessary and sufficient for the Hamiltonian description \cite{arnold}. In particular,
for any two functions $C(Q)$ and $D(Q)$
the Poisson bracket is defined as
\BEA
\label{pb}
\{C(Q), D(Q)\}_{PB}=\Omega^{-1}_{\,\, ab} \frac{\partial C}{\partial Q_a} \frac{\partial D}{\partial Q_b}.
\EEA
The equations of motion (\ref{buj}) are now written as
\BEA
\dot{Q}_a = \{\,Q_a, {\cal H}[Q]\,\}_{PB}.
\EEA
Note that the Poisson brackets are non-linear.
It is seen from (\ref{avgust}, \ref{pb}) that $z_{\a\b}$ and its
derivatives define a non-trivial symplectic structure for the system.

The matrix $Z$ is not invertible for an odd $K$. The Hamiltonian
description in this case is still possible, but requires more care in
explicitly accounting for constraints; compare with our discussion after
(\ref{barbarossa}).

\section{The fourth order Lagrangian.}
\label{+4}

Here we briefly report on the fourth-order Lagrangian. Since the
calculations now become very complicated, we shall restrict ourselves
to the situation where the classical system has just one single
coordiante $q$. For further simplicity we assume the quantum system has
real adiabatic eigenstates. In fact, the main purpose of this section is
to illustrate that the fourth-order Lagrangian again depends linearly on
the highest-order time-derivatives of the classical coordinate.

Following the same lines of calculation for the thrid order
non-adiabatic force presented in Appendix\ref{third}, and assuming a
single coordinate classical system and real eigenstates for the quantum
system, the non-adiabatic force acting on the classical system in the
fourth order is described by the following Lagrangian
\BEA\label{f4}
&&\eps^4 F^{[4]} = \left( \frac{d^3}{ds^3}\frac{\pd}{\pd q^{(3)}}  - \frac{d^2}{ds^2} \frac{\pd}{\pd \ddq}  + \frac{d}{ds} \frac{\pd}{\pd \dq} - \frac{\pd}{\pd q} \right) {\cal L}^{[4]}[q, \dq, \ddq, q^{(3)}],\\
&& {\cal L}^{[4]}[q, \dq, \ddq, q^{(3)}] = \eps^4 \left[a \dq^4 + b \ddq \dq^2 + w\dq   q^{(3)} \right],\label{L4}
\EEA
where $q^{(3)}$ stands for the third order time derivative of $q$,
and where ${\cal L}^{[4]}$ represents the fourth-order Lagrangian.
Note that the dependence on the higher-order time derivatives $\ddq$ and $q^{(3)}$ is linear.
The coefficients $a$, $b$, and $d$ are given as
\BEA
&&a(q) = \ssum'_k \frac{\vert \langle \partial_q N\vert k \rangle \vert^2}{\de_{nk}}
- \langle N \vert N \rangle \langle \partial_q n \vert N \rangle,
\label{a}\\
&& b(q) = -\ssum'_k
\frac{|\langle k|\partial_q n\rangle |^2}{\Delta^2_{nk}}\,\, \partial_q[\Delta_{nk}^{-1}]
,\label{b}\\
&& w(q)= - \ssum'_k \frac{|\langle k|\partial_q n\rangle |^2}{\Delta^3_{nk}}
\label{w},
\EEA
where $|N\rangle$ is given by (\ref{sochi}): $|N\rangle = \ssum_k' \frac{\langle k|\partial_q n\rangle }{\Delta_{nk}}|k\rangle$.

Then the total Lagrangian describing the one dimensional classical system reads
\BEA\label{L4tot}
{\cal L}_4[q, \dq, \ddq, q^{(3)}] = - V(q) - E_n(q)
+ \frac{\eps^2}{2}(M+G)\dq^2 + \eps^4\left( a \dq^4 + b \ddq \dq^2 + w\dq   q^{(3)}   \right),
\EEA
where the first and the third order terms vanish due to the assumption of real eigenstates of the quantum system, and where
$a$, $b$, and $w$ are given by (\ref{a})-(\ref{w}) and $G$ is defined as
\BEA\label{G}
G \equiv - 2 \ssum'_k\frac{ \langle n \vert \frac{d}{dq} \vert k \rangle ^2}{\de_{nk}}.
\EEA
The kinematics of this Lagrangian can be developed along the same lines as in the previous section.

\section{Summary}

We have studied the post-adiabatic equations of motion for a slow
classical system which is coupled to a fast quantum system. The slow
versus fast motion is controlled by a small ratio $\eps$ of the
characteristic times. The general problem we addressed is to find an
effective Lagrangian that describes the dynamics of the classical
system. The following facts were known for this problem: (1.) In the
order $\eps^0$ the effective Lagrangian differs from the bare one by the
Born-Oppenheimer potential energy. (2.) Berry and Robbins have shown
that the $\eps$-force term in the effective Lagrangian corresponds to a
magnetic field, which is related to the geometric phase \cite{robbins}.
(3.) Weigert and Littlejohn and Goldhaber has recently shown that in the
order $\eps^2$ the effective Lagrangian gets an additional kinetic
energy term, which is a second-order polynomial over the classical
velocities \cite{gold,lit}. 

In this work we obtain the following new results.

{\it i)} The post-adiabatic reaction force is proved to be Lagrangian up
to the fifth order in $\eps$. We conjecture that at every order of
$\eps$ the effective dynamics of the classical system can be derived
form a Lagrangian.

{\it ii)} Within the order $\eps^3$ the effective Lagrangian linearly
depends on the accelerations of the classical system.  

We argued that this result is important, because it provides a
physically well-motivated scenario for the emergence of
higher-derivative Lagrangians for open classical systems. This result
should be contrasted to the usual open-system approaches, which can also
produce forces depending on higher-order derivatives (e.g., the
Abraham-Lorentz force in electrodynamics), but those forces are
dissipative (non-Lagrangian). We also explained that this result does
not depend on the quantum character of the fast system, and will be
present as well if the fast system is classical and integrable [in the sense of \cite{arnold}].

The presence of higher-derivative terms can be tested by essential
influences they bring on the kinematics of the system.  First, they
modify initial conditions; in our case this means that the trajectory of
the classical system on the slow (coarse-grained) time starts to depend
on acceleration; see our discussion around (\ref{barsuk}--\ref{suksuk}). 
Second, the conserved energy of the slow classical
motion does depend on the acceleration. And third, the presence of
higher-derivative terms naturally separates the total angular momentum
into the sum of orbital momentum and spin.  We show that this spin
satisfies an exact analogue of the zitterbewegung relation. 

{\it iii)} We also found interesting results for the classical
autonomous dynamics within the order $\eps^2$, where the motion
generated by the effective classical Lagrangian can be mapped to
geodesic curves on a suitable Riemannian manifold. Operating with the
simplest possible example|two classical coordinates interacting with a
two-level quantum system|we show that the Riemannian manifold is
essentially curved solely due to the kinetic energy generated by the
fast quantum system. The scalar curvature is frequently negative
indicating that the classical trajectories [geodesic curves] are
unstable with respect to small variations of the initial conditions. The
metric tensor generated by the fast quantum system can change its
signature as a function of the two coordinates. Physically this means a
transition from an Euclidean to pseudo-Euclidean manifold, emergence of
a time-like coordinate and {\it etc}. This result deserves a careful
elaboration.

\subsection*{Acknowldegments}

We are grateful to Armen Nersissian for several discussions and for
explaining us the Hamiltonization procedure given in section
\ref{hamoo}. Ruben Manvelian is acknowledged for making several critical
remarks. Last but not least, we thank Theo Nieuwenhuizen for his kind
support.

\appendix

\section{On the derivation of the quantum-classical dynamics}
\label{apero}

The argument follows basically to \cite{bala,balian}. Let we are given a
two-degrees-of-freedom quantum system with Hamiltonian
$\frac{p^2}{2M}+V(q)+H(q,x)+\frac{\pi^2}{2M}$, where $q$ and $x$ are
operator coordinates, and where $p$ and $\pi$ are operator momenta. For
simplicity we shall assume that the initial state is factorized over
these two degrees of freedom. 

The Heisenberg equation generated by this Hamiltonian read:
\BEA
\label{buerak1}
&&\frac{\d q}{\d t}=\frac{p}{M},\\
\label{buerak2}
&&\frac{\d p}{\d t}=-V'(q)-H'_q(q,x),\\
\label{buerak3}
&&\frac{\d x}{\d t}=\frac{\pi}{m},\\
&&\frac{\d \pi}{\d t}=-H'_x(q,x).
\label{buerak4}
\EEA
Now the motion of the $(p,q)$ degree of freedom is separated into two part:
\BEA
\label{charles}
p(t)=\overline{p}(t)+p_f, \qquad q(t)=\overline{q}(t)+q_f, 
\EEA
where $\overline{p}(t)$ and $\overline{q}(t)$ are the averages over the initial state, and where $p_f$ and $x_f$ are the fluctuations.
By definition these operators satisfy 
\BEA
\label{cromwel}
\overline{p}_f=\overline{q}_f=0.
\EEA
We now substitute (\ref{charles}) into (\ref{buerak1}, \ref{buerak2}, \ref{buerak4}) and 
expand (\ref{buerak2}, \ref{buerak4}) over the small $q_f$:
\BEA
\label{buerak5}
&&\frac{\d \overline{q}}{\d t}+\frac{\d q_f}{\d t}=\frac{\overline{p}}{M}+\frac{p_f}{M},\\
\label{buerak6}
&&\frac{\d \overline{p}}{\d t}+\frac{\d p_f}{\d t}
=-V'(\overline{q})-H'_q(\overline{q},x)
-V''(\overline{q})q_f-H''_{qq}(\overline{q},x)q_f+{\cal O}(q_f^2),\\
&&\frac{\d \pi}{\d t}=-H'_x(\overline{q},x)-H''_{xq}(\overline{q},x)\, q_f+{\cal O}(q_f^2).
\label{buerak7}
\EEA
Averaging these equations over the initial state we obtain
\BEA
\label{buerak8}
&&\frac{\d \overline{q}}{\d t}=\frac{\overline{p}}{M},\\
\label{buerak9}
&&\frac{\d \overline{p}}{\d t}
=-\overline{V'(\overline{q})}-\overline{H'_q(\overline{q},x)}
-\overline{H''_{qq}(\overline{q},x)q_f}+\overline{{\cal O}(q_f^2)},\\
&&\overline{\frac{\d \pi}{\d t}}=-\overline{H'_x(\overline{q},x)}-\overline{H'_{xq}(\overline{q},x)\, q_f}
+\overline{{\cal O}(q_f^2)}.
\label{buerak10}
\EEA
If in (\ref{buerak9}, \ref{buerak10}) the terms proportional to ${\cal
O}(q_f)$ are neglected we get into a quantum-classical equations, where
$(\overline{p}, \overline{q})$ is the classical degree of freedom. If,
however, there are physical reasons to expect that the initial state
will remain factorized over the considered range of times [see
\cite{picard} for a detailed discussion of such situations], then we
have to neglect only terms ${\cal O}(q_f^2)$, because the terms ${\cal
O}(q_f)$ drop out due to (\ref{cromwel}). 

\section{Adiabatic Perturbation theory.}
\label{ap1}

Here we outline the adiabatic perturbation theory as developed in \cite{hage,baev}.
This appendix is completely self-contained and can be read independently from the
main text.

Consider the non-stationary Schr\"odinger equation
\begin{equation}
i \partial_t | \Psi(t)\rangle = H(\epsilon t) |\Psi(t)\rangle,
\label{a1}
\end{equation}
where $\hbar = 1$, $\eps\ll 1$ is a small parameter, and where $H(\epsilon t)$ is the slowly
changing Hamiltonian. Let us define
\begin{equation}\label{slowtime}
s = \epsilon t,
\end{equation}
for the slow time. Let $E_n(s)$ and $\vert n(s) \rangle$ be the adiabatic eigen-energies
and eigenvectors, respectively:
\begin{equation}
\label{eigenstate}
H(s) \vert n(s) \rangle = E_n(s) \vert n(s) \rangle, \qquad
\langle n(s) \vert m(s) \rangle = \delta_{nm}, \qquad n=1,\ldots,d.
\end{equation}
We assume that the adiabatic energy spectrum is non-degenerate for all $s$.
Let us also assume for simplicity that the system starts from the initial state equal to one
of the adiabatic eigen-states:
\BEA
| \Psi(0)\rangle = |n(0)\rangle.
\label{a5.0}
\EEA

The first step in any adiabatic approach is to separate the slowly
changing quantities from the fast ones. To this end we look for the solution of (\ref{a1}) as
\BEA
| \Psi(\epsilon,t)\rangle = |\psi_n(\epsilon,s)\rangle \, e^{i\alpha_n(t)}, \qquad
\alpha_n(t) \equiv - \int_0^t \d\hat{t} \, E_l(\epsilon \hat{t}) ,
\label{a5}
\EEA
where $\alpha_n(t)$ is the dynamical phase. It is clear that
$e^{i\alpha_n(t)}$ changes fast, i.e., as $\sim e^{i/\eps}$, for slow times $s$.
Putting (\ref{a5}) into (\ref{a1}) we get
\BEA
\label{a6}
 i \epsilon |\dot{ \psi}_n(\eps,s)\rangle  = \left[ H(s) - E_n(s)\right] |\psi_n(\eps,s)\rangle,
\EEA
where dot means differentiation over $s$.
Now we expand the slow wave-function, $\psi_n(\epsilon,s)$, over $\epsilon$
\begin{equation}
\label{a7}
\vert \psi_n(\epsilon,s) \rangle  =  e^{i \gamma_n(s)}\left[\, \vert n(s)\rangle +
\epsilon \vert n_1(s) \rangle + \epsilon^2 \vert n_2(s) \rangle + ... \right],
\end{equation}
where
\BEA
\gamma_n(s) = i \int_0^s \d u \,\langle n(s)\vert \dot n(u) \rangle ,
\EEA
is the Berry phase factor. We separated it out to facilitate further calculations.

Substituting power series expansion (\ref{a7}) into (\ref{a6}) and comparing
terms of equal order of $\epsilon$,
we get
\begin{eqnarray}
\label{a8}
 &&0 = \left( H - E_n \right)\, \vert n \rangle ,\\
\label{a9}
&&i \vert \dot n \rangle - i \langle n \vert \dot n \rangle \, \vert n \rangle = \left( H - E_n \right)\, \vert n_1 \rangle,\\
\label{a10}
&&i \vert \dot{n}_1 \rangle - i \langle n \vert \dot n \rangle \, \vert n_1 \rangle = \left( H - E_n \right)\, \vert n_2 \rangle,\\
&&\vdots\nonumber
\end{eqnarray}
Eq.~(\ref{a8}) holds automatically.
To solve the higher order equations we introduce the projection operators $P$ and $Q$:
\begin{eqnarray}\label{a11}
&&P = \vert n \rangle  \langle n \vert, \qquad
Q ={\ssum}'_k \vert k \rangle \langle k \vert,\\
&&P + Q = 1,\qquad PQ = QP = 0.
\end{eqnarray}
where ${\ssum}'_k$ means the term $k = n$ is excluded from the summation ${\ssum}_{k=1}^d$.

Operating $Q$ from left to the both sides of (\ref{a9}), we get
\begin{equation}
 i {\ssum_k}' \vert k \rangle  \langle k \vert \dot n \rangle
= {\ssum_k}' \Delta_{kn} \vert k \rangle  \langle k \vert n_1 \rangle,\qquad
\Delta_{kn} \equiv E_k - E_n.
\end{equation}
Since $\Delta_{k\not=n}$ is non-zero [energy levels are not degenerate], we get
\begin{equation}
\label{varaz}
\langle k \vert n_1 \rangle \,\equiv\,
c^{[1]}_{k\neq n} = - i \frac{\langle k \vert \dot n \rangle}{\Delta_{nk}}.
\end{equation}
This determines $Q \vert n_1 \rangle$, but we still have to find $P \vert n_1 \rangle$.
Let us define
\begin{eqnarray}
\label{a13}
&&  \vert n_1 \rangle = c^{[1]}_{nn} \, \vert n \rangle + \vert n_1^{\perp} \rangle, \\
&& \vert n_1^{\perp} \rangle \equiv {\ssum_k}' c^{[1]}_{kn} \, \vert k \rangle, \\
&& c^{[1]}_{nn} \equiv \langle n \vert n_1 \rangle,
\end{eqnarray}
where $c^{[1]}_{nn}$ has to be found.
To this end we multiply both sides of (\ref{a10}) from left by $P$
[recall that (\ref{a10}) is obtained from $\eps^2$ term in expansion (\ref{a7})]
\begin{equation}\label{a16}
 \langle n \vert \dot n_1 \rangle = \langle n \vert n_1 \rangle \, \langle n \vert \dot n\rangle,
\end{equation}
where $\vert \dot{n}_1 \rangle$ is found from (\ref{a13}):
\begin{equation}\label{Ap1n-dn1prim}
 \vert \dot{n}_1 \rangle = \dot{c}^{[1]}_{nn} \, \vert n \rangle
+ {c}^{[1]}_{nn}  \vert \dot n \rangle  +  \vert \dot{n}_1^{\perp} \rangle.
\end{equation}

Using (\ref{a16}) we get
\begin{equation}
\label{a17}
\dot{c}^{[1]}_{nn}(s)  = - \langle n(s) \vert  \dot{n}_1^{\perp}(s) \rangle,
\end{equation}
which together with (\ref{a5.0}) solves as
\BEA
\label{a18}
 c^{[1]}_{nn}(s) &=& - \int_0^s \, \d u \langle n(u) \vert  \dot{n}_1^{\perp} (u) \rangle
     =  - i {\ssum_k}' \int_0^s \d u\, \frac{|\langle k(u)| \dot{n}(u) \rangle|^2}{\Delta_{nk}(u)}.
\EEA
It is seen that $c^{[1]}_{nn}(s)$ is purely imaginary.

Analogous argument gives the higher order non-adiabatic corrections $\vert n_m(s) \rangle,
(m>1)$ in (\ref{a7})
\BEA
&& \vert n_m \rangle = {\ssum_k}' c^{[m]}_{kn} \, \vert k \rangle + c^{[m]}_{nn} \vert n \rangle, \\
&& c^{[m]}_{k\neq n} = \frac{i \langle n \vert\dot n \rangle \, c^{[m-1]}_{k\neq n}
- i \langle k \vert \dot{n}_{m-1} \rangle}{\Delta_{nk}},\\
&& c^{[m]}_{nn} = - {\ssum_k}' \int_0^s \d u\, c^{[m]}_{kn}(u) \, \langle n(u) \vert \dot{k}(u) \rangle .
\EEA

Altogether $\vert \psi_n(\epsilon,s) \rangle $ in (\ref{a7}) can be written as
\BEA
&&\vert \psi_n(\epsilon,s) \rangle =  e^{i \gamma_n(s)}\ssum_k c_{kn} \vert k \rangle,\\
&&c_{kn} = \delta_{kn} + \epsilon c^{[1]}_{kn} + {\epsilon}^2 c^{[2]}_{kn} +  \cdots.
\label{a22}
\EEA

Some relations between coefficients $c_{kn}$ can be uncovered without knowing their explicit form, but rather
looking at the normalization condition:
\BEA
\label{a20}
\ssum_k \vert c_{nk} \vert^2 = 1,
\EEA
which should be satisfied at each order of $\eps$. For the first two orders
\BEA
|c_{nn}|^2=1+2\eps\Re\{ c_{nn}^{[1]} \}+2\eps^2\Re\{ c_{nn}^{[2]} \}+\eps^2 | c_{nn}^{[1]}|^2
+{\cal O}(\eps^3),
\label{ab3}
\EEA
which brings in the following two relations
in the orders $\eps$ and $\eps^2$, respectively,
\BEA
\label{ab4}
&&   \Re\{ c_{nn}^{[1]} \}=0,\\
&&   2\Re\{ c_{nn}^{[2]} \}+| c_{nn}^{[1]}|^2+{\ssum_k}'|c_{kn}^{[1]}|^2=
2\Re\{ c_{nn}^{[2]} \}+\langle n_1|n_1\rangle=0.
\label{ab5}
\EEA

Let us work out explicitly the second-order coefficient $c^{[2]}_{k\not = n}(s)$.
Using in (\ref{a22}),
$\langle k|\dot{n}_1\rangle=\frac{d}{ds}[\, \langle k|n_1\rangle \,]-\langle \dot k|n_1\rangle$
we get
\BEA
\label{copu3}
c^{[2]}_{k\not =n}(s)=c^{[1]}_{k\not =n}(s)c^{[1]}_{nn}(s)+\frac{i}{\Delta_{nk}(s)}\left[
c^{[1]}_{k\not =n}\,\left(\,\langle n | \dot n \rangle-\langle k | \dot k \rangle\,\right)
-\frac{d}{ds}\left[\,c^{[1]}_{k\not =n}\,\right]+{\ssum_{l(\not=k)}}' \, c^{[1]}_{ln}\,\langle \dot k | l \rangle
\right].
\EEA

\subsection{Precision of the adiabatic approximation}

An important problem of precision of the adiabatic approximation was studied in \cite{lenard,super,hage}. Simplifying
previous results on this problem, Hagedorn and Joye have proven the following result \cite{hage}.
Following to (\ref{a7}) let us define
\BEA
\vert \psi^N(\epsilon,s) \rangle  =  e^{i \gamma_n(s)}\sum_{k=0}^N \epsilon^k \vert n_k(s) \rangle.
\EEA
Let $\{a\}$ define the integer part of a real number $a$, and let we are given a positive number $g$. Then it is shown that
\cite{hage}:
\BEA
\left|\,\,\vert \psi^{\{g/\epsilon  \}}(\epsilon,s) \rangle - \vert \psi_{\rm exact}(\epsilon,s) \rangle\,\,\right|
\leq C(g)e^{-\Gamma (g)/\epsilon},
\EEA
where $\vert \psi_{\rm exact}(\epsilon,s) \rangle$ is the exact solution
of the original equation (\ref{a6}) $\vert \psi^{\{g/\epsilon
\}}(\epsilon,0) \rangle=\vert \psi_{\rm exact}(\epsilon,0) \rangle$ at
the initial time, and where $C(g)$ and $\Gamma(g)$ are bounded positive
functions of $g$.  This results implies that the precision of the
adiabatic approximation is exponential over $\epsilon$. 


\section{Calculation of the third-order post-adiabatic force.}
\label{third}

Within the present Appendix $\ssum_k'$ means $\ssum_{k=1,\, k\not =n}^d$, and $\dot{A}=A'=\frac{\d A}{\d s}$, where $s$ is the
slow time.

The third-order post-adiabatic force is given by (\ref{g7}), where for $\langle \phi_n| H|\phi_n \rangle-E_n$ we should employ
(\ref{osiris}) and then (\ref{kora1}), while for $\Im \langle \pd_\mu \phi_n|\phi_n' \rangle $ we directly use (\ref{g3}).
Having done these, we select terms $\propto \eps^3$ and end up with
\BEA
\label{tartar1}
\frac{F^{[3]}}{2}=\pd_\mu {\ssum}_k' \Delta_{kn}\Re [c^{[2]}_{kn} c^{[1]*}_{kn}]
+\frac{\d}{\d s} \Im \langle \pd_\mu n|n_2 \rangle +
\pd_\mu \Im \langle n_2|n' \rangle +\Im \langle \pd_\mu n_1|n_1' \rangle \\
=\frac{\d}{\d s} \Im \langle \pd_\mu n|n_2 \rangle +
\pd_\mu \Im [c^{[2]*}_{nn} \langle n|n' \rangle] +\Im \langle \pd_\mu n_1|n_1' \rangle,
\label{tartar2}
\EEA
where we additionally
employed (\ref{varaz}) and $|n_2\rangle=c^{[2]}_{nn}|n\rangle+\ssum_k' c^{[2]}_{kn}|k\rangle$
when going from (\ref{tartar1}) to (\ref{tartar2}).

Looking at (\ref{copu3}) we introduce the following notation
\BEA
c^{[2]}_{k\not =n}=c^{[1]}_{k\not =n}c^{[1]}_{nn}+\tilde{c}^{[2]}_{k\not =n}.
\EEA

Using (\ref{ab5}) we exclude $\Re c^{[2]}_{nn}$ and get
\BEA
\label{tartar3}
\frac{F^{[3]}}{2}=-\Im\left[ \langle \pd_\mu n|n' \rangle\right] \,\langle n_1|n_1\rangle
-\frac{1}{2}\Im \left[\langle \pd_\mu n|n \rangle\right] \,\frac{\d}{\d s}\langle n_1|n_1\rangle
-\frac{1}{2}\Im \left[\langle  n|n' \rangle\right] \,\pd_\mu\langle n_1|n_1\rangle\\
+\Im \langle \pd_\mu n_1|n_1' \rangle
+\frac{\d}{\d s}\Im\left[
{\ssum}_k' {c}^{[2]}_{kn}\langle \,\pd_\mu n|k\rangle
\right].
\label{tartar4}
\EEA

Let us denote:
\BEA
|n_1\rangle=c_{nn}^{[1]}|n\rangle+|n_1^\bot\rangle.
\EEA
Using this notation and expanding $\Im \langle \pd_\mu n_1|n_1' \rangle$ we get after some algebraic steps
that all non-local terms in (\ref{tartar3}, \ref{tartar4}) are cancelled out. This means that
(\ref{tartar3}, \ref{tartar4}) are written as
\BEA
\label{tartar5}
 \frac{F^{[3]}}{2}=
-\Im\left[ \langle \pd_\mu n|n' \rangle\right] \,\langle n_1^\bot|n_1^\bot\rangle
-\frac{1}{2}\Im \left[\langle \pd_\mu n|n \rangle\right] \,\frac{\d}{\d s}\langle n_1^\bot|n_1^\bot\rangle
-\frac{1}{2}\Im \left[\langle  n|n' \rangle\right] \,\pd_\mu\langle n_1^\bot|n_1^\bot\rangle
\\
+\Im \langle \pd_\mu n_1^\bot |\frac{\d}{\d s}n_1^\bot \rangle
+\frac{\d}{\d s}\Im\left[
{\ssum}_k' \tilde{c}^{[2]}_{kn}\langle \,\pd_\mu n|k\rangle
\right].
\label{tartar6}
\EEA

Using (\ref{copu3}) (\ref{tartar5}, \ref{tartar6}) can be represented as
\BEA
\label{tartar7}
 \frac{F^{[3]}}{2}=
-\frac{1}{2}\frac{\d}{\d s}\left[
\Im\left( \langle \pd_\mu n|n \rangle\right) \,\langle n_1^\bot|n_1^\bot\rangle
\right]
-\frac{1}{2}\pd_\mu\left[
\Im \left(\langle n|n' \rangle\right) \langle n_1^\bot|n_1^\bot\rangle\right]
+\frac{\d}{\d s}\left[
\Im\left( \langle n|n' \rangle\right) \,\Re{\ssum}_k'
\frac{\langle \pd_\mu n|k\rangle\langle k|n'\rangle}{\Delta^2_{nk}}
\right]
\\
+\Im \langle \pd_\mu n_1^\bot |\frac{\d}{\d s}n_1^\bot \rangle
+\frac{\d}{\d s}\Im\left[
{\ssum}_k' \left(
-\frac{i}{\Delta_{nk}}\left[\,c^{[1]}_{kn}\,\right]'+\frac{i}{\Delta_{nk}}
{\ssum_{l}}' \, c^{[1]}_{ln}\,\langle k' | l \rangle
\right)\,
\langle \,\pd_\mu n|k\rangle
\right].
\label{tartar8}
\EEA
After some algebra one can show that (\ref{tartar7}) can be generated by a Lagrangian
\BEA
(\ref{tartar7})= \left( \frac{\d}{\d s}\frac{\pd}{\pd \dq_\mu}-\frac{\pd}{\pd q_\mu}  \right)
\frac{1}{3}h_{\a\b\g}\dq_\a\dq_\g\dq_\g,
\EEA
where
\BEA
h_{\a\b\g} &=& \frac{1}{2}\Im\left( \langle n|\pd_\g n \rangle\right) \,\Re[\langle N_\a|N_\b\rangle]
+
\frac{1}{2}\Im\left( \langle n|\pd_\a n \rangle\right) \, \Re[\langle N_\gamma|N_\b\rangle]
+\frac{1}{2}\Im\left( \langle n|\pd_\b n \rangle\right) \, \Re[\langle N_\a|N_\gamma\rangle].
\EEA
Here we defined
\BEA
|n_1^\bot\rangle=-i \dq_\a |N_\a\rangle, \qquad |N_\mu\rangle \equiv {\ssum}_k'
\frac{\langle k|\pd_\mu n\rangle}{\Delta_{nk}}\,|k\rangle,
\EEA
Recall that an implicit summation is carried over indices $\a,\b$ and $\g$.
Note that $h_{\a\b\g}$
is symmetric with respect to any permutation of indices $\a,\, \b$ and $\g$, so that
\BEA
\frac{1}{3}h_{\a\b\g}\dq_\a\dq_\g\dq_\g =
\frac{1}{2}\Im\left( \langle n|\pd_\g n \rangle\right) \,\Re[\langle N_\a|N_\b\rangle]\,
\dq_\a\dq_\g\dq_\g .
\EEA

We now work out (\ref{tartar8}):
Then
\BEA
\Im\left[
{\ssum}_k' \left(
-\frac{i}{\Delta_{nk}}\left[\,c^{[1]}_{kn}\,\right]'+\frac{i}{\Delta_{nk}}
{\ssum_{l}}' \, c^{[1]}_{ln}\,\langle k' | l \rangle
\right)\,
\langle \,\pd_\mu n|k\rangle
\right]\\
=\Im\left[
{\ssum}_k' \left(
\frac{1}{i\Delta_{nk}}\left[\,c^{[1]}_{kn}\,\right]'+\frac{1}{i\Delta_{nk}}
{\ssum_{l}}' \, c^{[1]}_{ln}\,\langle k | l' \rangle
\right)\,
\langle \,\pd_\mu n|k\rangle
\right]\\
=\Im \left[-i \langle N_\mu |\frac{\d}{\d s}n^\bot_1\rangle\right]=
-\Im \left[\langle N_\mu|N_\a\rangle \ddq_\a+\dq_\a\dq_\b \langle N_\mu | \pd_\b N_\a\rangle
\right],
\label{golo1}
\EEA
\BEA
\Im \langle \pd_\mu n_1^\bot |\frac{\d}{\d s}n_1^\bot \rangle =
\Im \left[\langle \pd_\mu N_\b|N_\a\rangle \ddq_\a\dq_\b+\dq_\a\dq_\b\dq_\g
\langle\pd_\mu N_\b | \pd_\g N_\a\rangle
\right]
\label{golo2}
\EEA

Combining together (\ref{golo1}, \ref{golo2}) we get

\BEA
(\ref{tartar8}) &=& \ddq_\a \dq_\b \Im\langle \pd_\mu N_\b|N_\a \rangle
+\dq_\a\dq_\b\dq_\g \Im \langle \pd_\mu N_\b|\pd_\g N_\a \rangle -\frac{\d}{\d s}\Im\left[
\langle N_\mu|N_\a\rangle \ddq_\a+\dq_\a\dq_\b \langle N_\mu |\pd_\b N_\a\rangle
\right]\\
\label{but1}
&=& \Im\langle N_\a|N_\mu \rangle\,\frac{\d^3}{\d s^3}q_\a
+\ddq_\a \dq_\b \Im\left[\,\pd_\b \langle  N_\a|N_\mu \rangle
+\langle \pd_\b N_\a|N_\mu \rangle +\langle \pd_\a N_\b|N_\mu \rangle
+\langle \pd_\mu N_\b|N_\a \rangle \,
\right]\\
&+& \dq_\a\dq_\b\dq_\g \Im
\left[\,\pd_\g \langle\pd_\b  N_\a|N_\mu \rangle
+\langle \pd_\mu N_\b|\pd_\g N_\a \rangle \,
\right]
\label{but2}
\EEA

We define
\BEA
z_{\a\b}\equiv \frac{1}{2}\Im \langle N_\b|N_\a \rangle
\EEA
and write
\BEA
\label{voron1}
&&(\ref{but1})+(\ref{but2})=
\left[
\frac{\d^2}{\d s^2}\frac{\pd  }{\pd \ddot{q}_\mu}
-\frac{\d}{\d s}\frac{\pd  }{\pd \dq_\mu}+\frac{\pd  }{\pd q_\mu}\right]
z_{\a\b}\ddq_\a \dq_\b\\
\label{voron2}
&&+\ddq_\a\dq_\b\,\,\frac{1}{2}\Im\left[\,
\langle \pd_\b N_\mu|N_\a \rangle+\langle \pd_\b N_\a|N_\mu \rangle
+\langle \pd_\a N_\mu|N_\b \rangle+\langle \pd_\a N_\b|N_\mu \rangle
+\langle \pd_\mu N_\a|N_\b \rangle+\langle \pd_\mu N_\b|N_\a \rangle\,
\right]\\
&&+\dq_\a\dq_\b\dq_\g\, \Im\left[\,
\pd_\g \langle \pd_\b N_\a | N_\mu \rangle +
\langle \pd_\mu N_\b | \pd_\g N_\a \rangle -\frac{1}{2}\pd^2_{\a\g}
\langle N_\b | N_\mu \rangle
\,\right],
\label{voron3}
\EEA
where in obtaining (\ref{voron2}) we employed:
\BEA
\Im \langle \pd_\b N_\mu|N_\a \rangle=\frac{1}{2}
\Im \langle \pd_\b N_\mu|N_\a \rangle+\frac{1}{2}
\Im \langle \pd_\b N_\mu|N_\a \rangle \\
= \frac{1}{2}
\Im \langle \pd_\b N_\mu|N_\a \rangle-\frac{1}{2}
\Im \langle N_\a|\pd_\b N_\mu \rangle\\
=
\frac{1}{2}\Im \langle \pd_\b N_\mu|N_\a \rangle
+\frac{1}{2}\Im \langle \pd_\b N_\a| N_\mu \rangle
+\frac{1}{2} \pd_\b\Im \langle N_\mu| N_\a \rangle.
\EEA
The quantity inside of the square brackets in (\ref{voron2}) is symmetric with respect to
permutation of indices $\mu$, $\a$ and $\b$. We now
define
\BEA
&&\lambda_{\a\b\g}=
\frac{1}{4}\Im\left[\,
\langle \pd_\b N_\g|N_\a \rangle+\langle \pd_\b N_\a|N_\g \rangle
+\langle \pd_\a N_\g|N_\b \rangle+\langle \pd_\a N_\b|N_\g \rangle
+\langle \pd_\g N_\a|N_\b \rangle+\langle \pd_\g N_\b|N_\a \rangle\,
\right]\\
&&\lambda_{\a\b\g}\dq_\a\dq_\b\dq_\g=\dq_\a\dq_\b\dq_\g\,\frac{6}{4}\Im\left[\,
\langle \pd_\a N_\b|N_\g \rangle\,
\right]
\EEA
and obtain
\BEA
(\ref{voron2})+(\ref{voron3}) &=&
\left[\frac{\d}{\d s}\frac{\pd  }{\pd \dq_\mu}-\frac{\pd  }{\pd q_\mu}\right]
\frac{1}{3}\lambda_{\a\b\g}\\
\label{ca2}
&+&\dq_\a\dq_\b\dq_\g\,\Im\left[\,
\pd_\g \langle \pd_\b N_\a | N_\mu \rangle +
\langle \pd_\mu N_\b | \pd_\g N_\a \rangle -\frac{1}{2}\pd^2_{\a\g}
\langle N_\b | N_\mu \rangle\right.\\
&+&\left.
\frac{1}{2}\pd_\mu\langle\pd_\a  N_\b | N_\g \rangle
-\frac{1}{2}\pd_\g\langle\pd_\a  N_\mu | N_\b \rangle
-\frac{1}{2}\pd_\g\langle\pd_\b  N_\a | N_\mu \rangle
-\frac{1}{2}\pd_\g\langle\pd_\mu  N_\a | N_\b \rangle
\,\right]
\label{ca3}
\EEA

Working out (\ref{ca2}, \ref{ca3}) we finally obtain:
\BEA
(\ref{ca2})+( \ref{ca3})=0.
\EEA

Thus the third-order post-adiabatic force is purely Lagrangian (though containing
higher-order derivatives):
\BEA
\frac{F^{[3]}}{2}= \left( \frac{\d}{\d s}\frac{\pd}{\pd \dq_\mu}-\frac{\d^2}{\d s^2}\frac{\pd}{\pd \ddq_\mu}
-\frac{\pd}{\pd q_\mu}  \right)
\left[\frac{1}{3}(h_{\a\b\g}+\lambda_{\a\b\g})\dq_\a\dq_\g\dq_\g
-z_{\a\b}\ddq_\a\dq_\b\right].
\EEA

\end{document}